\documentclass{aa}

\usepackage{txfonts}
\usepackage{graphicx}
\usepackage{subfig}
\usepackage{natbib}
\bibpunct{(}{)}{;}{a}{}{,}

\begin{document}

\title{Forced turbulence in thermally bistable gas: A parameter study}

\author{D. Seifried\inst{1,3} \and W. Schmidt\inst{2,3} \and J.C. Niemeyer\inst{2,3}}

\institute{Institut f\"ur Theoretische Astrophysik, Universit\"at Heidelberg, Albert-Ueberle-Str. 2, 69120 Heidelberg, Germany\\
           \email{dseifried@ita.uni-heidelberg.de}
\and Institut f\"ur Astrophysik G\"ottingen, Universit\"at G\"ottingen, Friedrich-Hund-Platz 1, 37077 G\"ottingen, Germany
\and Lehrstuhl f\"ur Astronomie, Institut f\"ur theoretische Physik und Astrophysik, Universit\"at W\"urzburg, Am Hubland, 97074 W\"urzburg, Germany}

\date{Received / Accepted }

\abstract{Thermal instability is one of the dynamical agents for turbulence in the diffuse interstellar medium, where both turbulence and thermal instability interact in a highly non-linear manner.}
{We study basic properties of turbulence in thermally bistable gas for variable simulation parameters.  The resulting cold gas fractions can be applied as parameterisation in simulations on galactic scales.}
{Turbulent flow is induced on large scales by means of compressive stochastic forcing in a periodic box. The compressible Euler equations with constant UV heating and a parameterised cooling function are solved on uniform grids. We investigate several values of the mean density of the gas and different magnitudes of the forcing. For comparison with other numerical studies, solenoidal forcing is applied as well.}
{After a transient phase, we observe that a state of statistically stationary turbulence is approached. Compressive forcing generally produces a two-phase medium, with a decreasing fraction of cold gas for increasing forcing strength. This behaviour can be explained on the basis of turbulent mixing. We also find power-law tails of probability density functions of the gas density in high-resolution runs. Solenoidal forcing, on the other hand, appears to prevent the evolution into a two-phase-medium for certain parameter regions.}
{The dynamics of thermally bistable turbulence show a substantial sensitivity to the initial state and the forcing properties.}

\keywords{hydrodynamics - instabilities - turbulence - methods:numerical - ISM:kinematics and dynamics}

\maketitle

\section{Introduction}

Observations of the interstellar medium (ISM) show large variations in the thermodynamical properties of the gas, which led to the famous three-phase-model of \citet{MO77}. In this article, we focus on the warm neutral medium (WNM) and the cold neutral medium (CNM), whose occurrence was first explained consistently by \citet{Field65} with the help of the thermal instability (TI). The classical conception of a more or less static multiphase medium with different phases coexisting in pressure equilibrium \citep[see, for instance,][]{Wolfire95} has nowadays been replaced by a picture of a highly turbulent ISM \citep[e.~g.][]{BallKless07,KeeOst07}. Several observational results point to turbulence in the ISM, for example, the broadening of spectral lines, chemical mixing of the ISM, or the star formation rate \citep[for an overview, see][]{MacLow04,ElmeScalo04}. As turbulence decays roughly on one dynamical timescale \citep[e.~g.][]{MacLow98,Stone98}, energy has to be injected into the ISM to maintain turbulent motion. Several different sources were proposed to drive turbulence in the ISM, e.~g., blast waves from supernova explosions, the differential rotation of galaxies, but also thermal or gravitational instabilities.

While the properties of forced turbulence in the molecular and nearly isothermal phase have been studied extensively \citep[e.~g.][]{Kritsuk07,Schmidt09,FederDuv09}, there is no corresponding analysis for the diffusive ISM so far. Previous work on turbulence in thermally bistable gas encompass decaying turbulence \citep{Kritsuk02}, simulations on kpc scales to study the effect of stellar feedback \citep{Joung06,Joung09}, and colliding flows of atomic gas subject to the thermal instability \citep{Heitsch06,Vazquez07,Hennebelle07a,Banerjee09,AuditHenne09,FoliniWald09}. The last scenario has attracted a lot of attention as a possible model for the formation of molecular clouds from the diffuse ISM. Furthermore, simulations of turbulence produced by purely solenoidal forcing in thermally bistable gas were performed by \citet{Vazquez03}, \citet{Gazol05}, \citet{Kissmann08}, and \citet{Gazol09}. In these simulations, large amounts of thermally unstable gas (up to 50\%) were found, which is also indicated by observational studies \citep[e.~g.][]{Dickey77,Spitzer95,Heiles01,Heiles03,Kanekar03}.

As was demonstrated by \citet{Schmidt09} for the isothermal case, compressive stochastic forcing produces a flow structure on large scales that resembles colliding flows, because converging shocks are formed between rarefied gas regions. In this work, we study the properties of compressively driven turbulence in thermally bistable gas, depending on the magnitude of the force and the initial state of the gas, which is given by the equilibrium temperature for the mean gas density. In this regard, our approach is similar to the work of \citet{Gazol05}. The mean density is constant because of the periodic boundary conditions. The cooling function follows the definition of \citet{Audit05}. As \citet{Fed08} reported a substantial sensitivity of the density field in isothermal supersonic turbulence on the forcing \citep[see also][]{FederDuv09,Fed09}, we applied solenoidal forcing as well. Because of the three-dimensional parameter space (mean density, magnitude, and ratio of solenoidal to compressive modes of the forcing), the aim of the present study is to investigate the general behaviour of forced thermally bistable turbulence by means of a series of moderate-resolution simulations, using uniform grids with $256^3$ cells. To avoid a further increase in dimensionality of the parameter space, we omit magnetic fields, self-gravity, and thermal conduction. Nevertheless, our results serve as a useful guide estimating the influence of unresolved thermal processes in galactic disk simulations \citep{TaskBry06,DobbsGlov08,AgerLake09}. In such simulations, the cutoff length at the highest level of refinement is typically of the same order of magnitude as the forcing length scale in the periodic-box simulations. In the various prescriptions of the local star formation efficiency, it is neglected that the fraction of cold gas that can turn into stars via gravitational collapse is less than unity. We therefore investigate the influence of the forcing and the mean density on this cold gas fraction in detail. Future work will concentrate on particular parameter sets to obtain the turbulent scaling and fragmentation properties from high-resolution simulations and to study the influence of additional physics.

The outline of this article is as follows. Section \ref{sc:numerics} briefly describes the numerical techniques. The parameters, for which we run simulations, are specified in Sect. \ref{sec:simulations}. The phenomenology of turbulence in thermally bistable gas is described for a representative parameter set in Sect. \ref{sec:results}. For this parameter set, the grid resolution was varied from $128^3$ to $512^3$ to check for the resolution dependency of the results. Next, we analyse the variation of statistical properties for different simulation parameters. In Sect. \ref{sec:discussion}, the results are discussed in the context of other numerical studies and observations, followed by a summary of the main results and an outlook in Sect. \ref{sec:conclusion}.

\section{Numerical techniques}
\label{sc:numerics}

The simulation presented in this article are performed with the open source code Enzo \citep{OShea05}. We solve the compressible Euler equations using the piecewise-parabolic method (PPM) of \citet{Colella84}. Besides an external driving force \textbf{\textit{f}}, which is responsible for generating turbulent motions, we include different heating and cooling terms, combined into a net cooling rate per unit mass ${\fam=2 L}$. The resulting equations for the mass density $\rho$, the velocity $\vec{u}$, and the specific total energy $e$ of the fluid can be written as
\begin{eqnarray}
\frac{D}{Dt} \rho &=& -\rho \nabla \cdot \vec{u} \,, \label{equ:cont} \\
\rho \frac{D}{Dt} \vec{u} &=& -\nabla P + \rho \vec{f} \label{equ:momentum} \,, \\
\rho \frac{D}{Dt}e + \nabla \cdot P\vec{u} &=& \rho \vec{f} \cdot \vec{u} - \rho {\fam=2 L}(\rho,\mathcal{T})\,,
\label{equ:energy}
\end{eqnarray}
where the total time derivative is defined by
\begin{equation}
\frac{D}{Dt} = \frac{\partial}{\partial t} + \vec{u} \cdot \nabla\,.
\end{equation}
The specific total energy $e$ is given by
\begin{equation}
e = \frac{1}{2} u^2 + \frac{P}{(\gamma - 1)\rho}\,,
\end{equation}
where $\gamma = 5/3$ is the adiabatic exponent of monatomic gas, and the gas pressure $P$ is related to the mass density and the temperature $\mathcal{T}$ via the ideal gas law:
\begin{equation}
P = \frac{\rho k_B \mathcal{T}}{\mu m_H}\,. \label{equ:igl}
\end{equation}
The constants $k_B$, $\mu$, and $m_H$ denote the Boltzmann constant, the mean molecular weight, and the mass of a hydrogen atom, respectively. 

The cooling function $\rho{\fam=2 L}$ was defined by \citet{Audit05}. It includes the fine-structure cooling of \ion{C}{ii} and \ion{O}{i}, as well as the cooling by H (Ly$\alpha$--line) and by electron recombination onto charged grains. The only heating process considered is the photoelectric effect on small grains and polycyclic aromatic hydrocarbons (PAH) caused by far-ultraviolet galactic background radiation. For detailed information about the different processes see \citet{Wolfire95, Wolfire03, Spitzer78} and \citet{Bakes94}. The cooling is implemented in the form of an explicit scheme into Enzo \citep{Niklaus09}. After each time step, the state variables are updated in several subcycles by adding the resulting total energy increment. As the considered cooling and heating processes are only well defined in the diffuse ISM, the calculation of ${\fam=2 L}$($\rho$,$\mathcal{T}$) is constrained to temperatures $\mathcal{T}$ $\in$ [10 K, 10\,000 K]. For gas with temperatures above \mbox{10\,000 K}, we use ${\fam=2 L}$($\rho$,10\,000 K), which leads to an underestimate of the emitted energy, hence to warmer gas than expected. In strongly driven simulations we apply a temperature cutoff at \mbox{30\,000 K}, assuming that the gas would cool down very fast beyond this limit. Since the speed of sound increases with the temperature, the cutoff allows us to avoid very small time steps arising from unphysical values of the temperature outside the range of the cooling function. The concerned simulation runs are mentioned in Sect. \ref{sec:results}.

The specific force $\vec{f}$ in Eqs. (\ref{equ:momentum}) and (\ref{equ:energy}), which generates large-scale motions of the gas, is composed in Fourier space. Each mode of the force field is given by a stochastic process that is based on the Ornstein-Uhlenbeck process \citep{Eswaran88,Schmidt06}. For the details of this method, see \citet{Schmidt09}. The forcing varies on the autocorrelation timescale $T$, which is given by the ratio of the integral length $L$ to the characteristic velocity $V$ of the large-scale motions. We set the integral length to half of the box size. Thus, the wavelengths of the forcing modes range from $L/2$ to $2L$. The free parameter $V$ specifies the magnitude of the forcing, i.~e., the specific force is $\sim V^{2}/L$. By applying a Helmholtz decomposition in Fourier space, the ratio of solenoidal (divergence-free) to compressive (rotation-free) modes can be arbitrarily adjusted. We use the decomposition parameter $\zeta$ such that $\zeta=1$ corresponds to purely solenoidal forcing, and the force becomes increasingly compressive as $\zeta$ decreases from $1$ to $0$.

\section{Simulation parameters} \label{sec:simulations}

Since we are interested in the local structure of the diffuse ISM, we treat the simulation domain as a part of a larger region with -- in the statistical sense -- homogeneous properties. Numerically, this is done by using a cubic simulation domain with periodic boundaries and a linear size of \mbox{$X = 40$ pc.} At the simulation start, the box is filled with atomic hydrogen of uniform density $\rho_{0}$ and temperature $\mathcal{T}_{0}$. For the mean particle densities $n_0$ (which remains constant because of the periodic boundaries), we chose \mbox{1.0 cm$^{-3}$}, \mbox{1.8 cm$^{-3}$} and \mbox{3.0 cm$^{-3}$}, which are comparable to the mean density of the diffuse ISM. In the initial phase, the nearly homogeneous gas will adjust very fast to the equilibrium temperature that is given by the condition ${\fam=2 L}=0$, regardless of the choice of $\mathcal{T}_{0}$. For this reason, we set $\mathcal{T}_{0}$ equal to the equilibrium temperature corresponding to $n_0$.

\begin{figure}
\resizebox{\hsize}{!}{\includegraphics{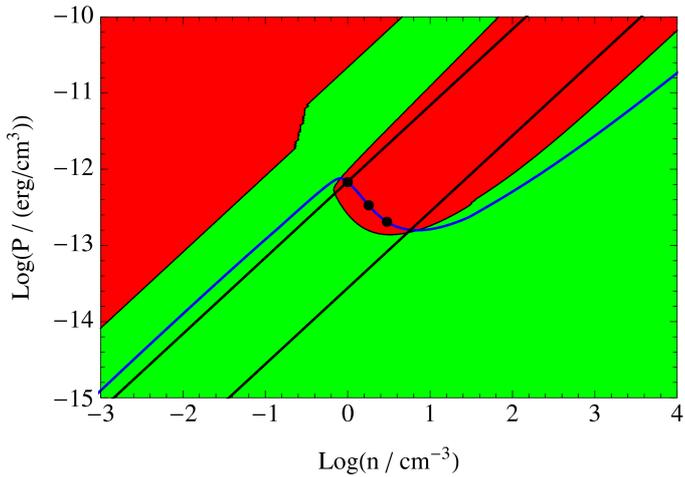}}
\caption{Pressure equilibrium curve for the used cooling and heating processes. Via the isobaric instability criterion \citep{Hunter70} the regions of thermal instability (red) and thermal stability (green) are determined. In addition, the isothermal lines at \mbox{$\mathcal{T} = 200$ K} and \mbox{$\mathcal{T} = 5000$ K} are shown, which define our region of thermally unstable gas. The dots represent the starting points of the simulations, which all lay in the thermally unstable region.}
\label{fig:1}
\end{figure}

The pressure equilibrium curve $P_{\mathrm{eq}}(n)$ following from \mbox{${\fam=2 L}=0$} and the ideal gas equation (see Sect.~\ref{sc:numerics}), is plotted in Fig. \ref{fig:1}.
Via the isobaric instability criterion of \citet{Hunter70}, one can determine the regions of the thermal instability (TI) in the phase diagram that is shown in Fig.~\ref{fig:1}. We define three different phases, a cold phase with \mbox{$\mathcal{T} <$ 200 K}, an unstable phase with \mbox{200 K $\leq \mathcal{T} <$ 5000 K}, and a warm phase with \mbox{$\mathcal{T} \geq$ 5000 K}. This classification has also been used in other works \citep{Heiles01,Heiles03,Audit05} and is warranted by the positions of the isothermal lines at \mbox{$\mathcal{T}$ = 200 K, 5000 K}, which -- of course only very roughly -- describe the region of TI in the intermediate and high-density range. In the following, the terms cold, unstable, and warm phase always refer to the above definitions. The chosen initial conditions of our simulations are located in the thermally unstable region (Fig. \ref{fig:1}). In consequence, one can expect that a two-phase-medium will evolve. As explained in Sect. \ref{sc:numerics}, we start the simulations with the gas at rest, and turbulence is generated by stochastic forcing. The forcing strength is regulated by the characteristic velocity $V$. Normalising $V$ by the initial soundspeed $c_0 = (\gamma k_B \mathcal{T}_0/\mu m_H)^{1/2}$, we obtain a characteristic Mach number, $\mathrm{Ma} = V/c_0$, which determines the magnitude of the stochastic force. We performed simulations with Ma between 0.2 and 3.0. However, Ma is not an indicator for the average Mach number of the flow, which depends on the phase separation of the gas. 

In one set of simulations, we applied purely compressive forcing ($\zeta$ = 0) for different initial densities. In addition, simulations with \mbox{$n_0$ = 1.0 cm$^{-3}$} were also carried out for the solenoidal case ($\zeta = 1$). The parameters of all simulations are listed in Table \ref{table:sim}.
\begin{table}
\caption{Performed simulations with the mean density $n_0$, the starting temperature $\mathcal{T}_0$, the Mach number Ma, $\zeta$ and the simulation time, normalised to the dynamical timescale $T$.}
\label{table:sim}
\centering
\begin{tabular}{c c c c c c}
\hline\hline
Run & $n_0$ (cm$^{-3}$) & $\mathcal{T}_0$ (K) & Ma & $\zeta$ & t ($T$)\\
\hline
D1Ma0.2 & 1.0 & 4908 & 0.25 & 0 & 6.0 \\
D1Ma0.5 & 1.0 & 4908 & 0.5 & 0 & 7.5 \\
D1Ma1.0 & 1.0 & 4908 & 1.0 & 0 & 6.0 \\
D1Ma1.5 & 1.0 & 4908 & 1.5 & 0 & 5.0 \\
D1Ma2.0 & 1.0 & 4908 & 2.0 & 0 & 5.0 \\
D1Ma1.0\_128 & 1.0 & 4908 & 1.0 & 0 & 6.0 \\
D1Ma1.0\_512 & 1.0 & 4908 & 1.0 & 0 & 6.0 \\
\hline
D1.8Ma0.2 & 1.8 & 1353 & 0.2 & 0 & 4.0 \\
D1.8Ma0.5 & 1.8 & 1353 & 0.5 & 0 & 5.0 \\
D1.8Ma1.0 & 1.8 & 1353 & 1.0 & 0 & 5.0 \\
D1.8Ma2.0 & 1.8 & 1353 & 2.0 & 0 & 5.0 \\
D1.8Ma3.0 & 1.8 & 1353 & 3.0 & 0 & 4.0 \\
\hline
D3Ma0.5 & 3.0 & 493 & 0.5 & 0 & 3.0 \\
D3Ma1.0 & 3.0 & 493 & 1.0 & 0 & 5.0 \\
D3Ma2.0 & 3.0 & 493 & 2.0 & 0 & 5.0 \\
D3Ma3.0 & 3.0 & 493 & 3.0 & 0 & 5.0 \\
\hline
D1Ma0.5s & 1.0 & 4908 & 0.5 & 1 & 3.0 \\
D1Ma1.0s & 1.0 & 4908 & 1.0 & 1 & 5.0 \\
D1Ma2.0s & 1.0 & 4908 & 2.0 & 1 & 5.0 \\
\hline
\end{tabular}
\end{table}
Each simulation ran until an approximate statistically steady state was found for at least one dynamical timescale $T$. Although it is not obvious that a well-defined statistically steady state can be reached for non-isothermal turbulence, all simulations presented here show a convergence of statistical quantities related to the flow properties and the thermodynamic properties of the gas as time proceeds. Thus, the statistics described in the next sections are averaged over the last dynamical timescale. Since the computational cost of performing simulations of turbulence in non-isothermal gas is significantly higher than for isothermal gas, all runs were performed on a static grid of $N = 256^3$ cells. The only exceptions are the runs D1Ma1.0\_128 with a resolution of 128$^3$ and D1Ma1.0\_512 with 512$^3$, to test the resolution dependency of the results. Because of the moderate resolution of the simulations, we only investigated global statistics and probability density functions, but not two-point statistics or power spectra.

\section{Results}
\label{sec:results}

\begin{figure*}
\centering
\subfloat[D1Ma1.0\_128]{\includegraphics[width=88mm]{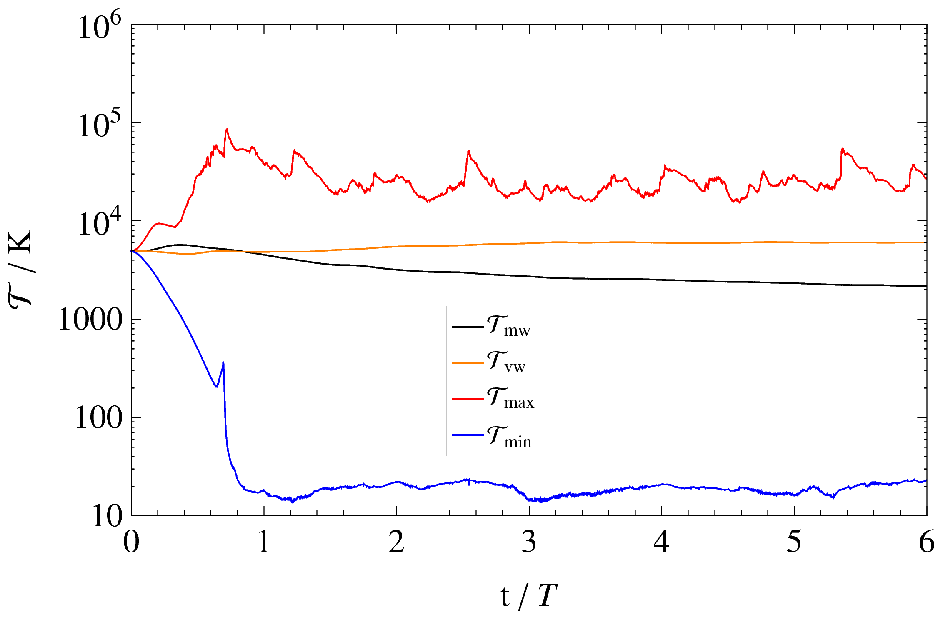}}
\subfloat[D1Ma1.0\_128]{\includegraphics[width=88mm]{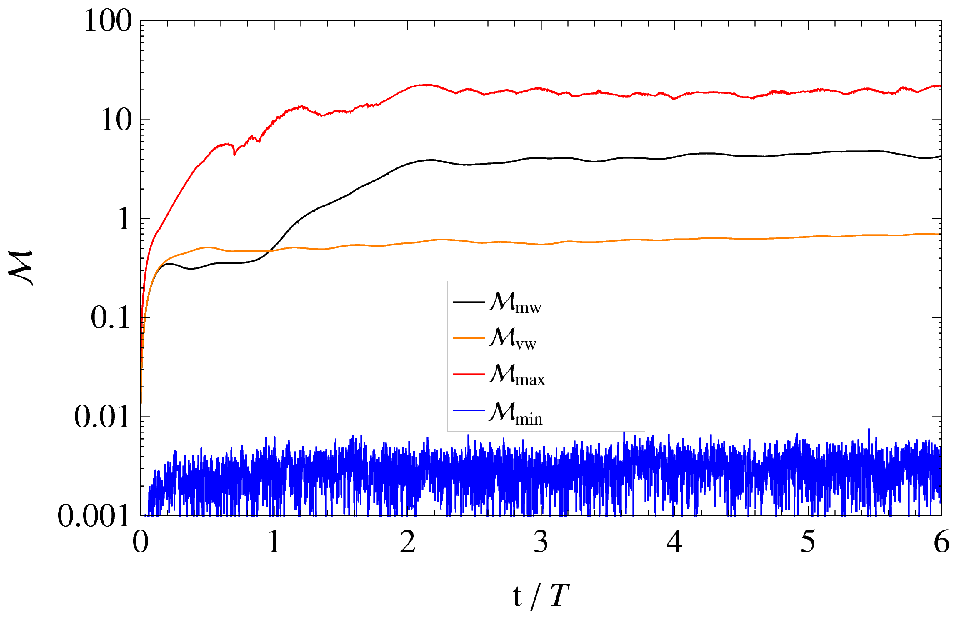}}\\
\subfloat[D1Ma1.0]{\includegraphics[width=88mm]{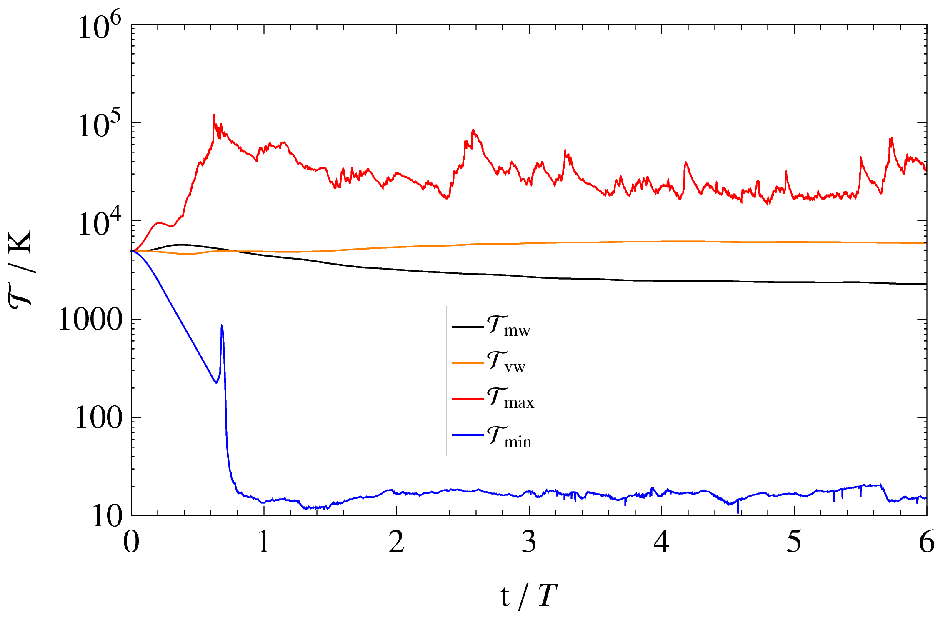}}
\subfloat[D1Ma1.0]{\includegraphics[width=88mm]{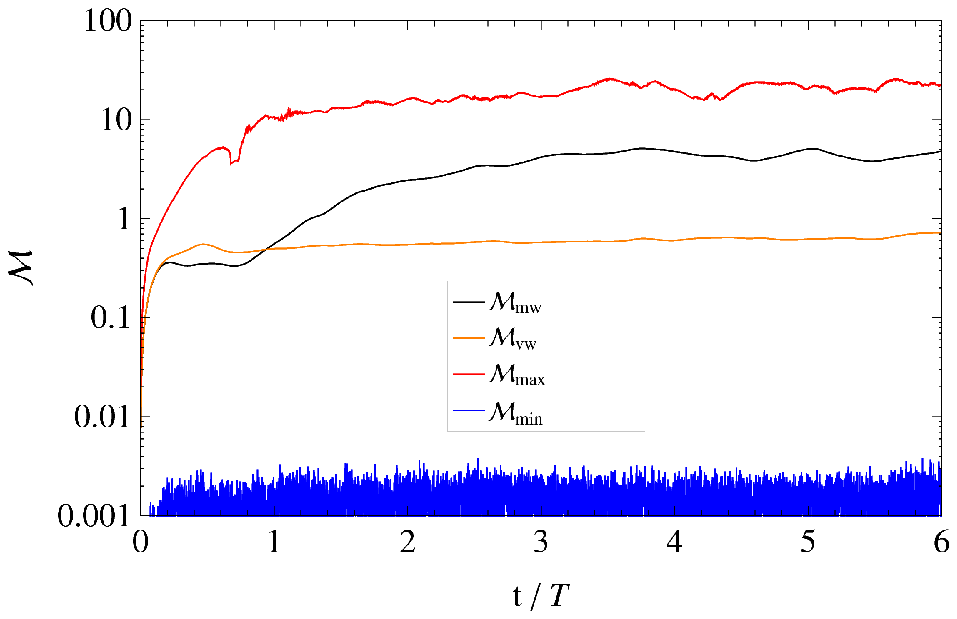}}\\
\subfloat[D1Ma1.0\_512]{\includegraphics[width=88mm]{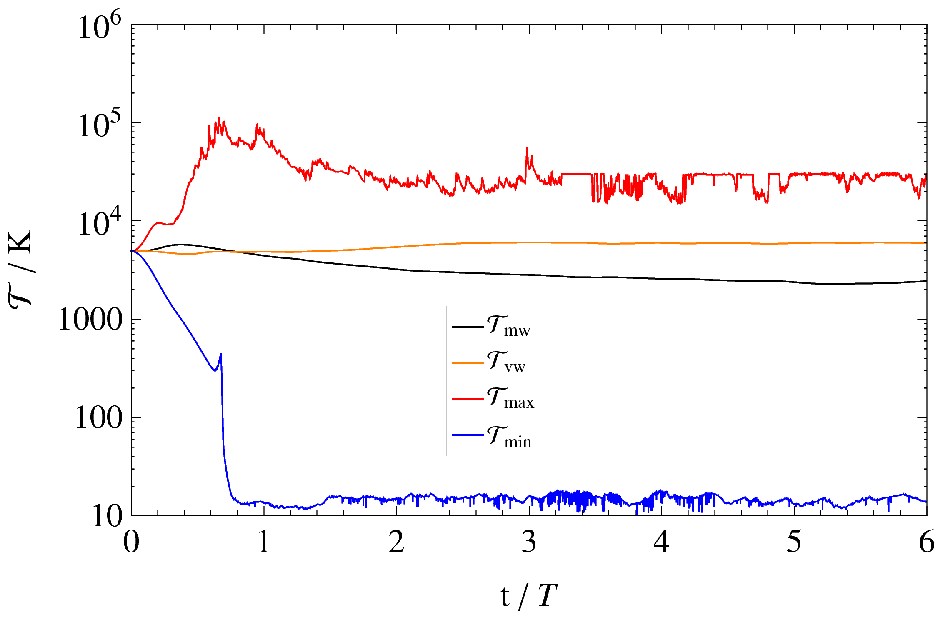}}
\subfloat[D1Ma1.0\_512]{\includegraphics[width=88mm]{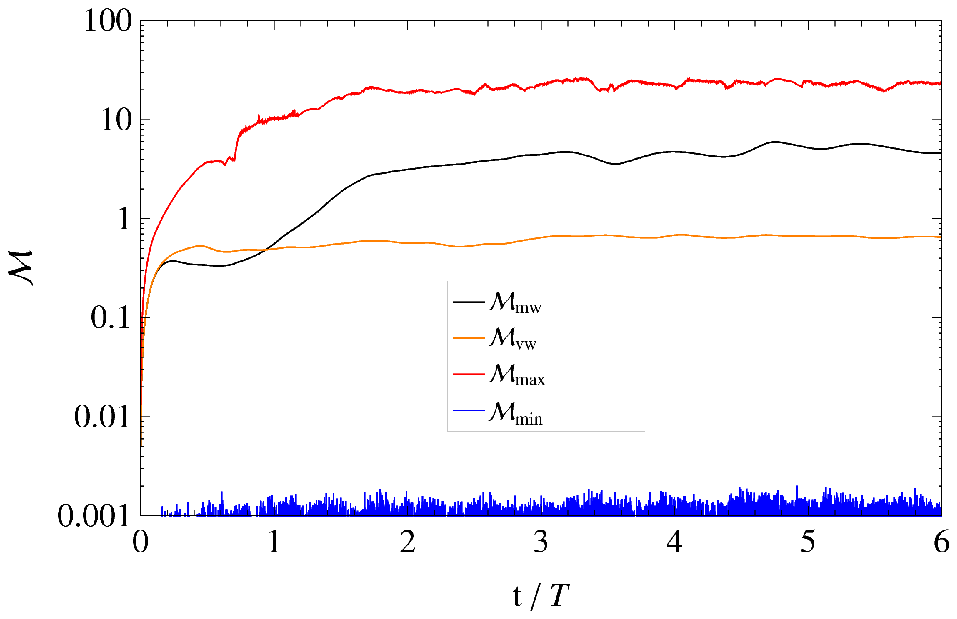}}
\caption{Time evolution of the temperature field (left panel) and the Mach number (right panel) for the simulation runs D1Ma1.0\_128 (upper panel), D1Ma1.0 (mid panel), and D1Ma1.0\_512 (lower panel). Plotted are the volume-weighted mean values, the mass-weighted mean values, and the maximum and minimum values.}
\label{fig:2}
\end{figure*}

\begin{figure*}
\centering
\includegraphics[width=88mm]{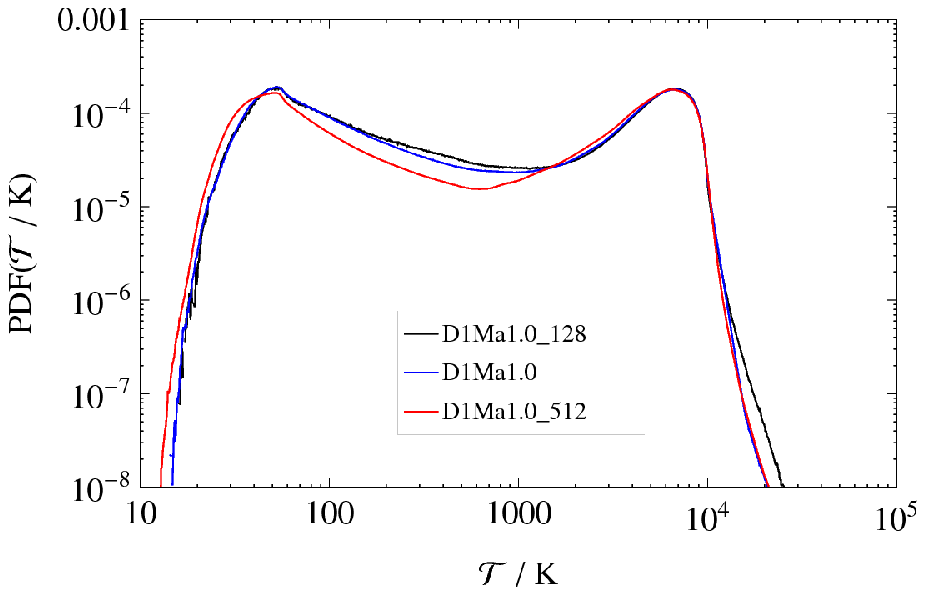}\hspace{12pt}
\includegraphics[width=88mm]{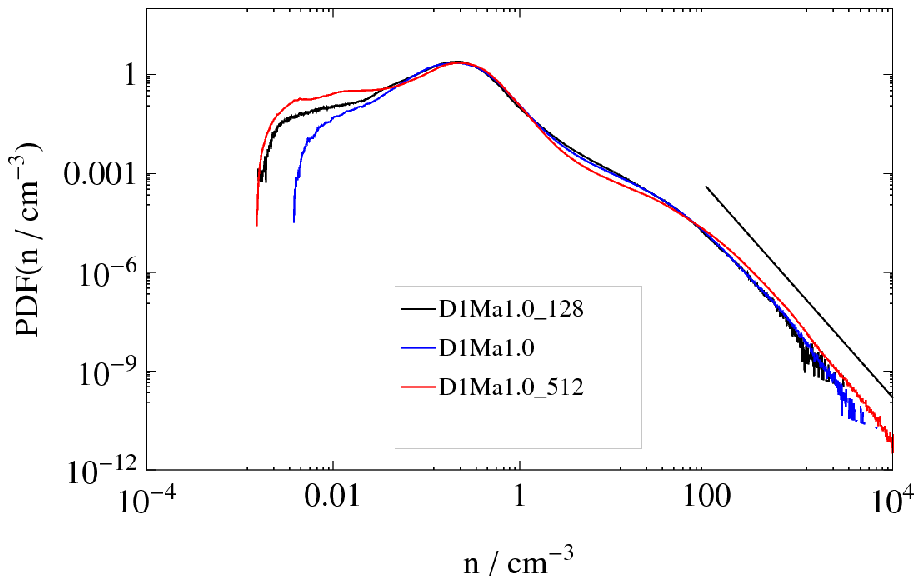}
\caption{Left: Temperature PDFs for the runs D1Ma1.0\_128, D1Ma1.0 and D1Ma1.0\_512, averaged over the last dynamical timescale. The maxima around 55 K and 6500 K indicate the CNM and the WNM. Right: Density PDFs for the same runs as in the left panel. For comparison a power law with an exponent of -3.2 is shown.}
\label{fig:3}
\end{figure*}

\begin{figure*}[t]
\centering
\includegraphics[width=80mm]{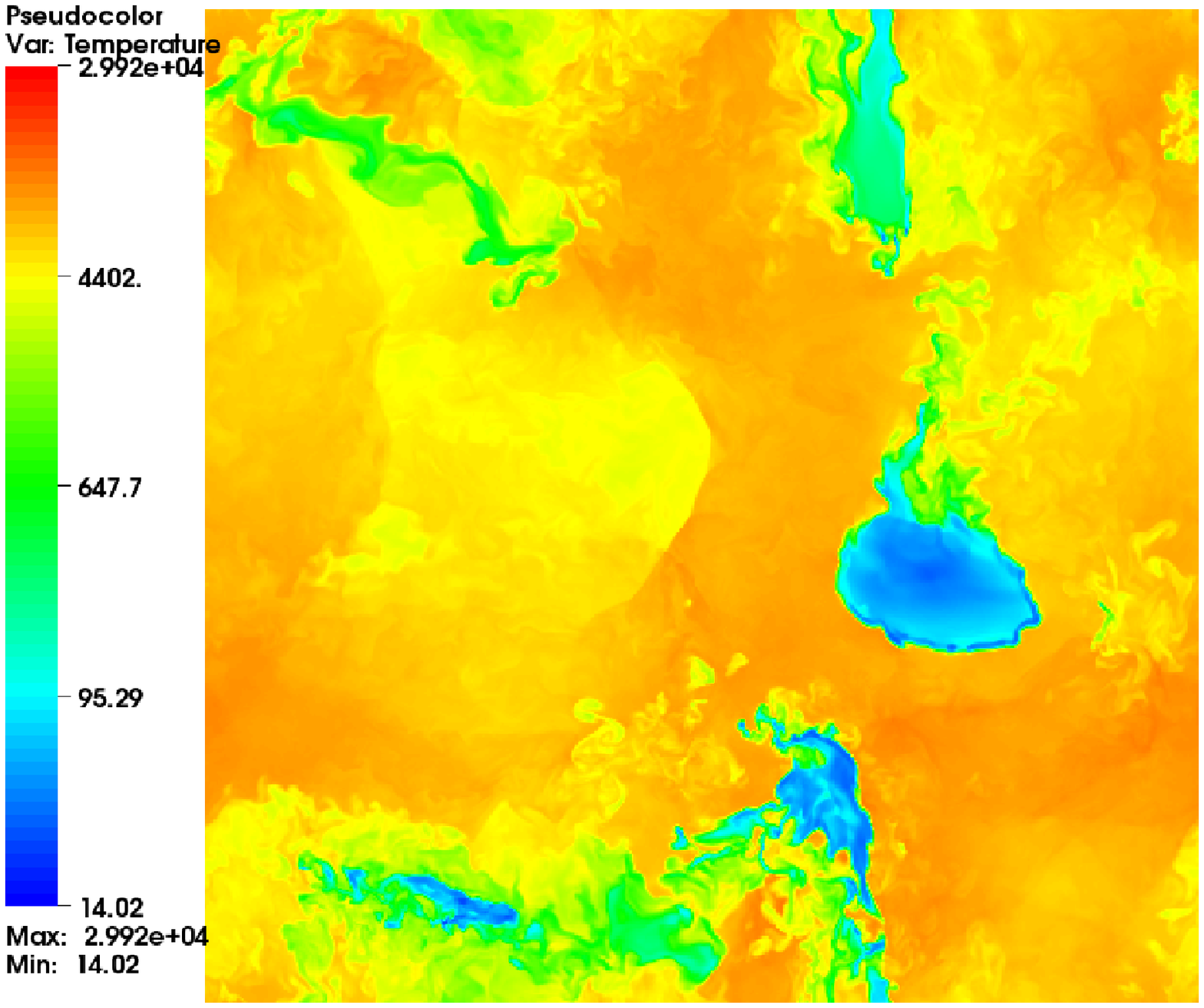}\hspace{12pt}
\includegraphics[width=80mm]{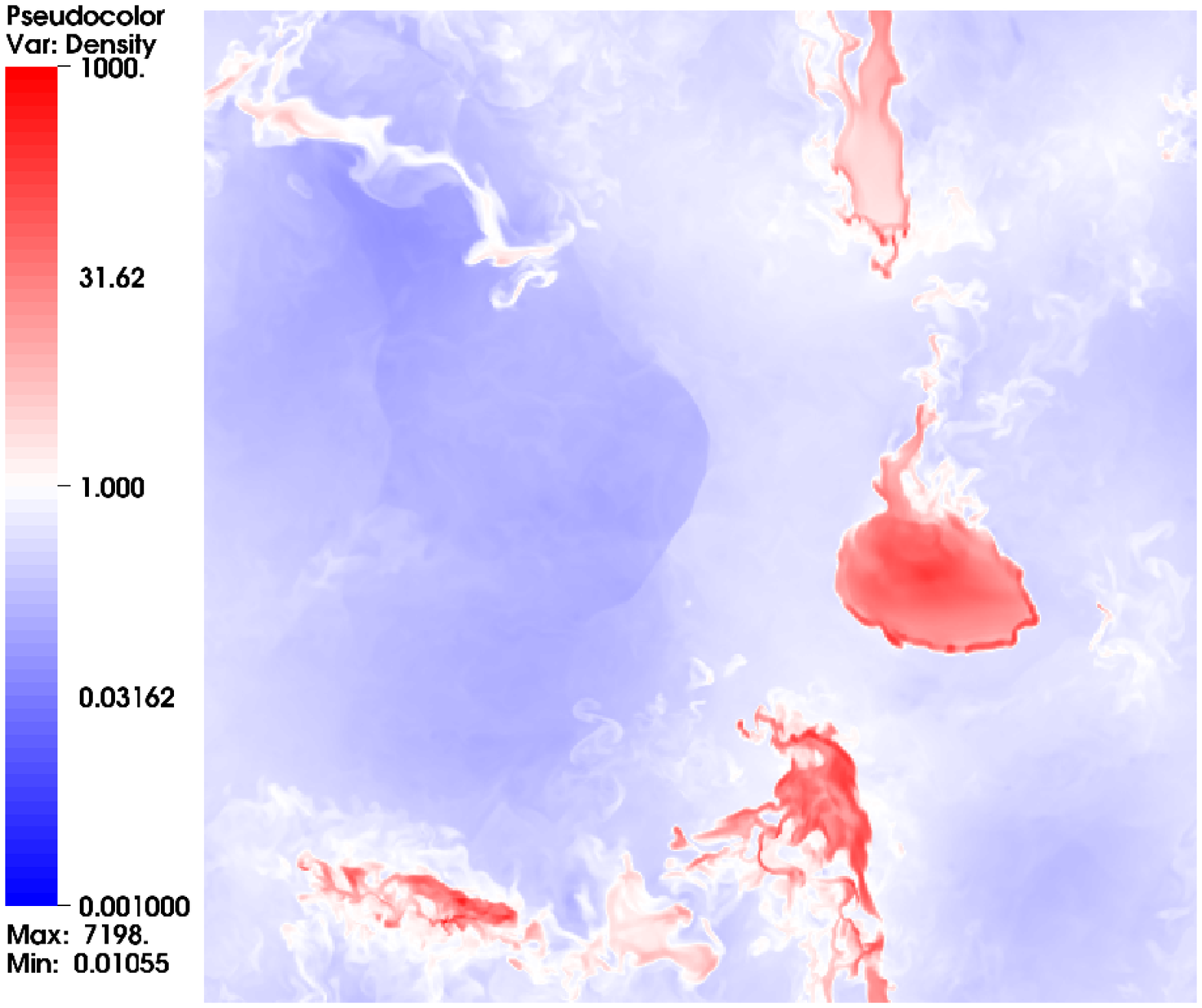}\\[12pt]
\includegraphics[width=80mm]{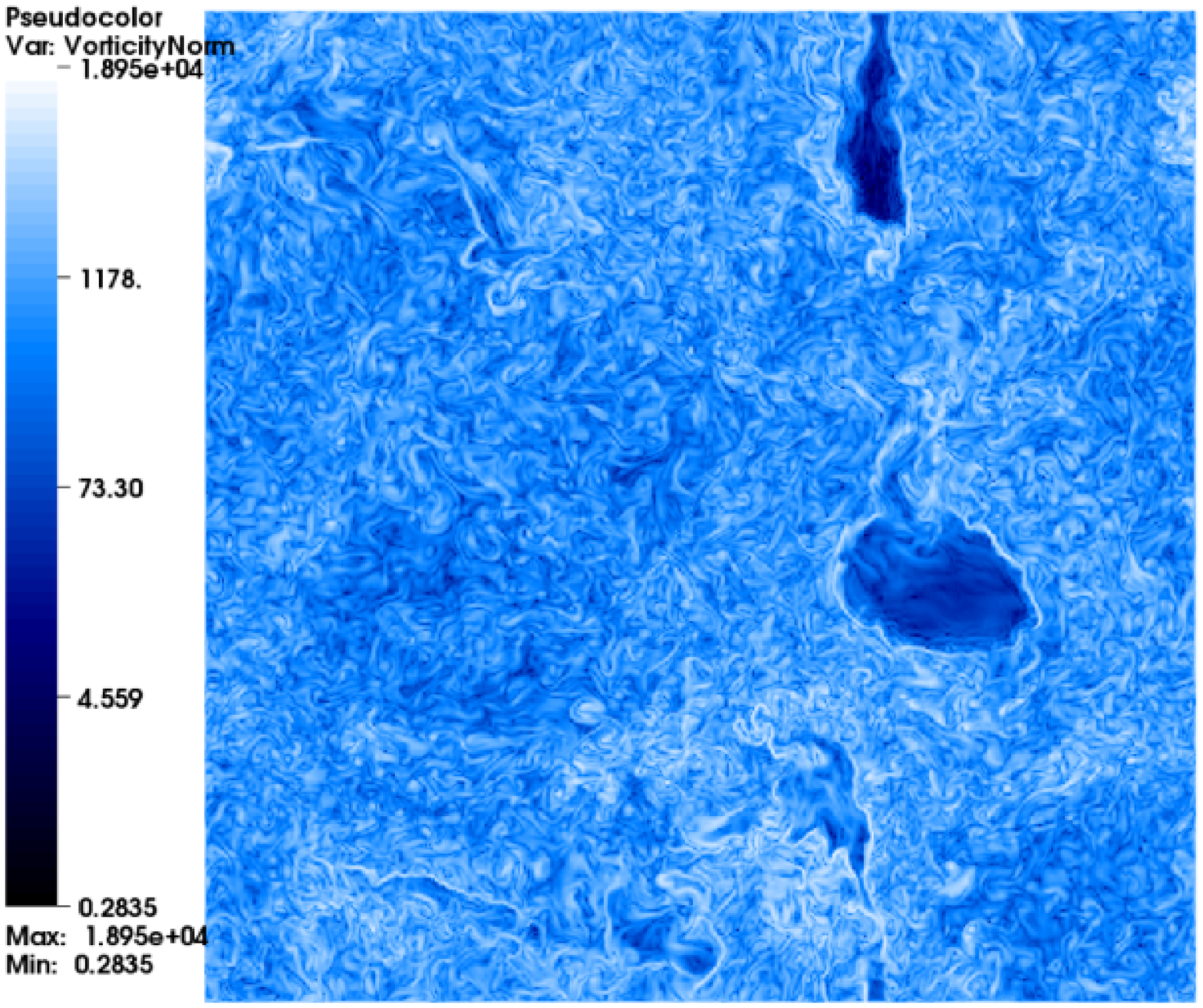}\hspace{12pt}
\includegraphics[width=80mm]{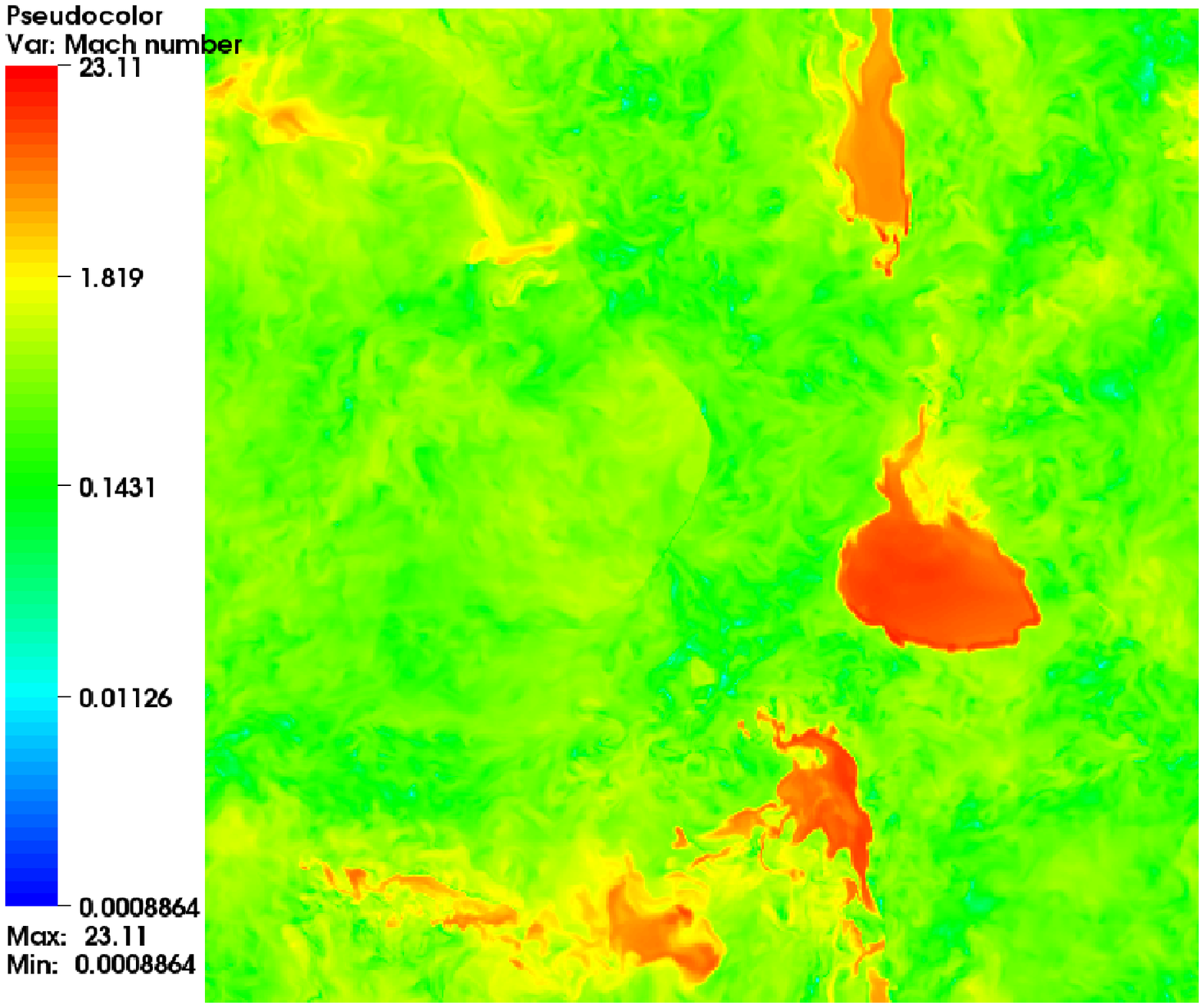}
\caption{Temperature (top left), density (top right), vorticity field (bottom left), and Mach number (bottom right) in a slice through the simulation domain of run D1Ma1.0\_512 at \mbox{t = 6.0 $T$}.}
\label{fig:4}
\end{figure*}

\subsection{Basic properties and resolution study} \label{sec:resstudy}

A crucial question for the following analysis is to what extent statistical properties are affected by the numerical resolution. To investigate the dependence on resolution, we consider the simulations D1Ma1.0\_128, D1Ma1.0 and D1Ma1.0\_512 with compressive forcing, a mean number density $n_{0}=1.0\,\mathrm{cm^{-3}}$, and $\mathrm{Ma=1.0}$. The resolution varies from $128^{3}$, over $256^{3}$ to $512^{3}$. Figure \ref{fig:2} shows the time evolution of the temperature field and the Mach number for these three simulations. For them, we see a similar evolution in time both for the temperature field and the velocity field (represented here by the Mach number). The onset of the TI becomes apparent by a rapidly spreading temperature distribution within the first dynamical timescale, where the temperatures finally range from roughly \mbox{10 K} to several \mbox{10\,000 K.} For the lower resolutions, we did not apply the cutoff at \mbox{30\,000 K}. As one can see, the temperature rises locally to as much as $10^{5}\ \mathrm{K}$. For the simulation with $512^3$ grid cells, these extrema entailed problems with the time stepping. Consequently, we introduced the high-temperature cutoff for $t > 3$ $T$. After a few dynamical timescales a statistically steady state is approached, in which the mass- and volume-weighted mean values are nearly constant. As can be seen in Fig. \ref{fig:2}, the mass-weighted mean of the temperature, $\mathcal{T}_{\mathrm{mw}}$, is much lower than the volume-weighted temperature $\mathcal{T}_{\mathrm{vw}}$, because the cold gas is associated with high-mass densities. The low temperature of the cold, dense gas results in a supersonic mass-weighted mean Mach number, $\mathcal{M}_{\mathrm{mw}}$, while $\mathcal{M}_{\mathrm{vw}}$ is transonic.

Averaging the different mean values over the last dynamical timescale, $\langle \rangle_t$, we only find small differences between the time-averaged mean values for the different resolutions. For the mass-weighted temperature mean values, which correspond to the mean thermal gas energy, we obtain \mbox{$\langle \mathcal{T}_{\mathrm{mw}} \rangle_t$ = 2229 K}, \mbox{2333 K}, and \mbox{2326 K} for simulation D1Ma1.0\_128, D1Ma1.0, and D1Ma1.0\_512, respectively. Thus, the relative deviations are less than 5\%. For the volume-weighted values \mbox{$\langle \mathcal{T}_{\mathrm{vw}} \rangle_t$}, the deviations are even less than 0.6\%. For \mbox{$\langle \mathcal{M}_{\mathrm{vw}}  \rangle_t$} and \mbox{$\langle \mathcal{M}_{\mathrm{vw}} \rangle_t$}, on the other hand, the differences are below 3.7\% and 20\%, respectively. The larger deviations of the time-averaged mean Mach numbers point toward a stronger sensitivity of the turbulent velocity field on the numerical resolution, while the thermodynamic properties appear to be quite robust.

Further indicators for the resolution dependency are probability density functions (PDFs). The PDFs of the number density and the temperature, averaged over the last dynamical timescale in the simulations D1Ma1.0\_128, D1Ma1.0, and D1Ma1.0\_512, are plotted in Fig. \ref{fig:3}. The two maxima of the temperature PDFs show that a cold and a warm gas phase are formed, with some gas in the unstable phase in between. The positions of the two maxima around \mbox{55 K} and \mbox{6500 K} are almost independent of the numerical resolution. These temperatures can be associated with the CNM and the WNM. Near the extrema of the temperature, there are slight deviations between the PDFs for different resolutions. The small excess of low-temperature gas in run D1Ma1.0\_512 corresponds to a slightly higher amount of very dense gas. The tiny fraction of gas hotter than \mbox{10\,000 K} supports our assumption that the cutoff at \mbox{30\,000 K} in run D1Ma1.0\_512 has no significant impact on the dynamics of the whole system. In the temperature interval from roughly \mbox{100 K} up to \mbox{1000 K}, we see a decrease in the amount of gas as the resolution increases. Consequently, the phase separation becomes more pronounced at higher resolution. The density PDFs shown in the right panel of Fig. \ref{fig:3} reveal a small excess of high-density gas for simulation D1Ma1.0\_512, which corresponds to the aforementioned increase in cold gas. For the highest resolution, the density PDFs also show a decrease in the amount of gas in the intermediate density range, which corresponds to thermally unstable gas. A notable feature in the high-density range is the power-law shape, which extends over more than one order of magnitude. We estimate the power-law exponent of run D1Ma1.0\_512 to roughly -3.2, whereas no power-law exponents are calculated for the lower resolution runs. Nevertheless, the fit for run D1Ma1.0\_512 seems to apply quite well even for run D1Ma1.0\_128 and D1Ma1.0, as shown in the right panel of Fig. \ref{fig:3}. The power law at these high densities is caused by an effective equation of state (EOS) in this range, which shows a polytropic index $\gamma_{\mathrm{eff}} < 1$. 

Although the statistics we investigated indicate that the simulations D1Ma1.0\_128 and D1Ma1.0 are not fully converged with respect to the density and temperature distributions, we conclude that the approximate behaviour of thermally bistable turbulence can be inferred from simulations with $256^3$ grid cells. This is the aim of the parameter study that is presented in the following.

\begin{figure*}[t]
\centering
\includegraphics[width=80mm]{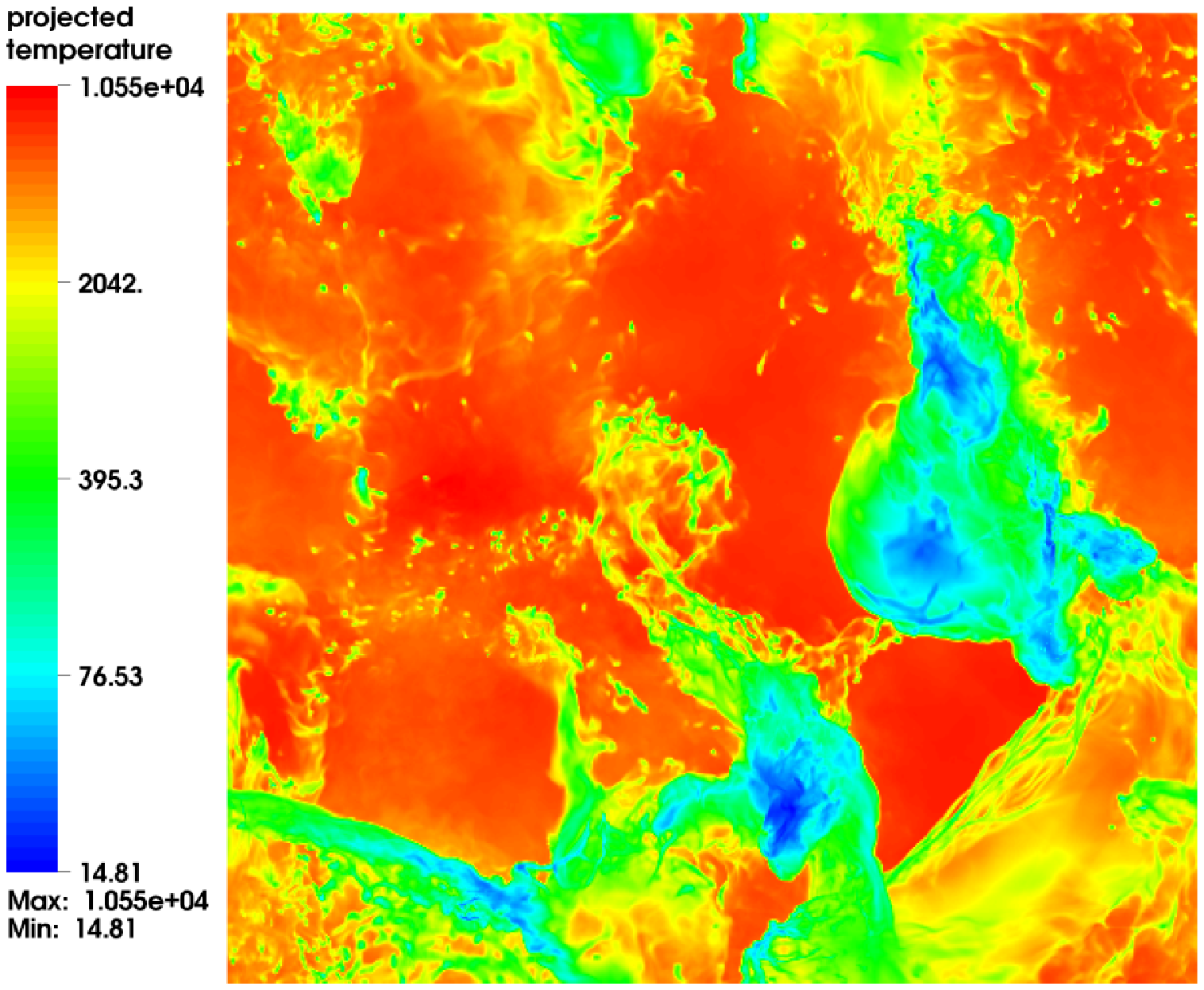}\hspace{12pt}
\includegraphics[width=80mm]{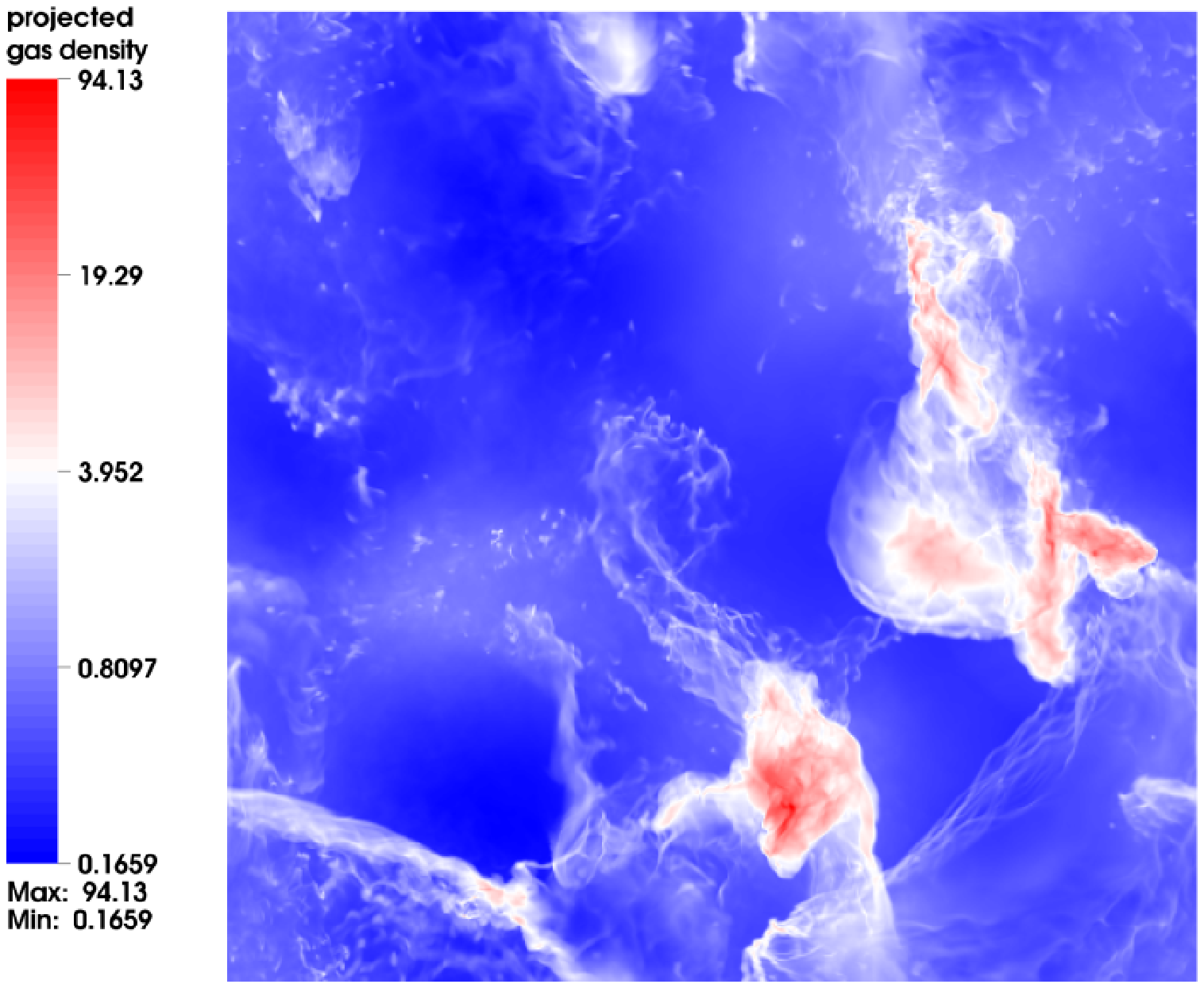}
\caption{Projected temperature (left) and density (right) fields for run D1Ma1.0\_512 at \mbox{$t$ = 6.0 $T$}.}
\label{fig:5}
\end{figure*}

\begin{figure}[h]
\resizebox{\hsize}{!}{\includegraphics{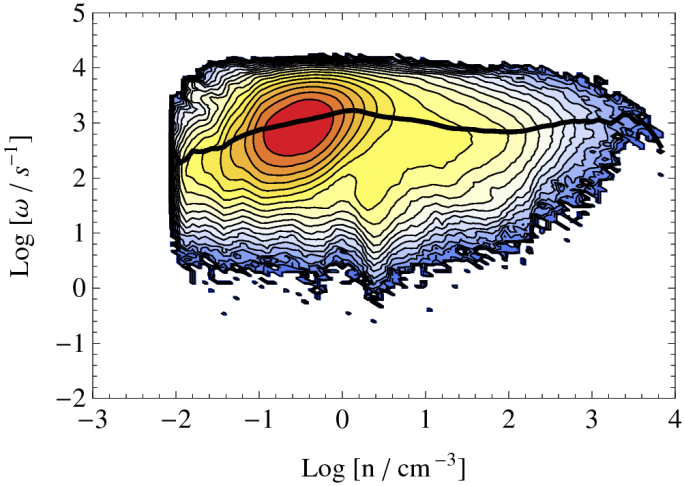}}
\caption{Two-dimensional PDF of the vorticity vs.\ the density for run D1Ma1.0\_512 at \mbox{$t$ = 6.0 $T$} with logarithmic scaled contour lines. The thick black line shows the density-dependent mean vorticity.}
\label{fig:6}
\end{figure}

To illustrate the properties of thermally bistable gas, Fig. \ref{fig:4} shows contour plots of the particle density, the temperature, the vorticity norm and the Mach number in 2D slices for simulation D1Ma1.0\_512 in the statistically stationary regime. When comparing the slices in the top panel of Fig. \ref{fig:4}, one can see that regions of high gas density correspond to low temperatures and vice versa. The most prominent feature that can be seen in the 2D slices is the big clump of cold, dense gas. While the boundaries of this clump are rather sharp, there are also regions in which the cold and warm gas phases entrain each other and form intricate structures. The rich small-scale structure becomes apparent in projections of the mass density and the temperature, which are shown in Fig. \ref{fig:5}. Even in these projections, relatively large clumps of cold gas appear prominently. Remarkably, the vorticity is largely reduced in these clumps in comparison to the surrounding medium, which is filled by more or less homogeneous turbulence (see left bottom panel of Fig. \ref{fig:4}). Nevertheless, the two-dimensional PDF of the vorticity magnitude versus the density, which is shown in Fig. \ref{fig:6}, demonstrates that high vorticity is found for the whole range of gas densities. Also the density-dependend mean value shows no clear trend but is more or less constant over the whole range of densities. This seems to contradict the observation made in the vorticity slice in Fig. \ref{fig:4}. A possible explanation could be that the vorticity remains low only in larger clumps of overdense gas, while in smaller clumps or in thin filamentary, but nevertheless overdense structures, the vorticity is about as high as in the surrounding low-density gas. Presumably, turbulent eddies that are produced by the TI in the unstable phase rapidly decay inside compact cold gas regions, because of the short sound crossing time. The compressive forcing, on the other hand, does not directly generate vorticity and it fails to induce shocks inside the clumps of extremely dense gas. It is possible, however, that these clumps would fragment if the numerical resolution were increased. The Mach number, on the other hand, indicates supersonic motions in the cold, dense gas and subsonic flow in the WNM (see right bottom panel of Fig. \ref{fig:4}). This leads to the effect that the mass weighted mean value $\mathcal{M}_{\mathrm{mw}}$ is higher than the volume weighted mean value $\mathcal{M}_{\mathrm{vw}}$ as shown in the right panel of Fig. \ref{fig:2}.

\begin{figure*}[t]
\centering
\includegraphics[width=80mm]{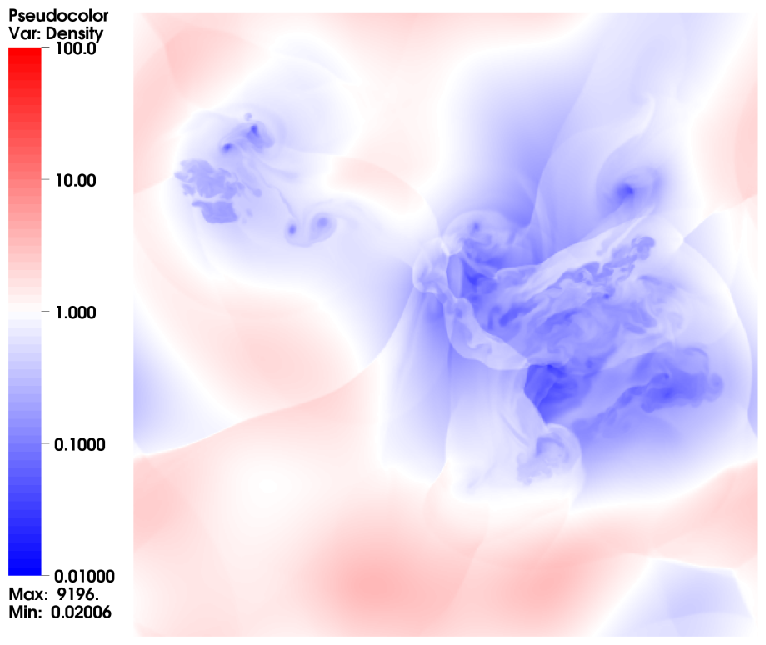}\hspace{12pt}
\includegraphics[width=80mm]{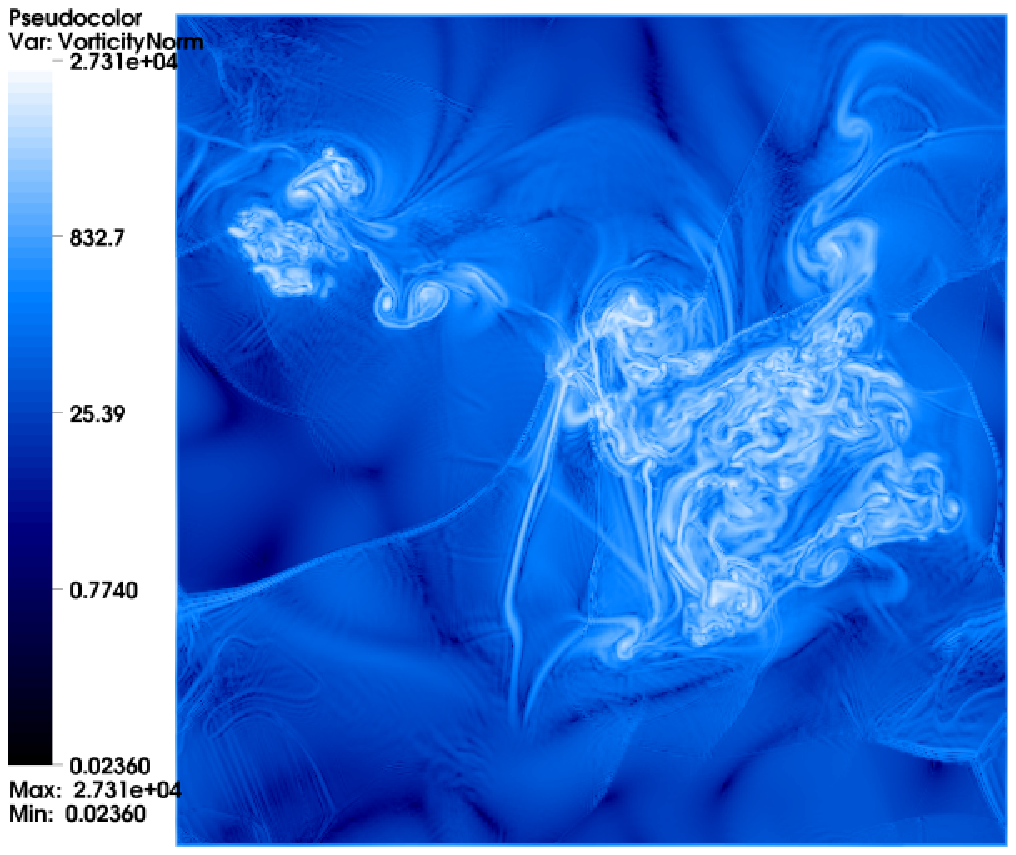}\\[12pt]
\includegraphics[width=80mm]{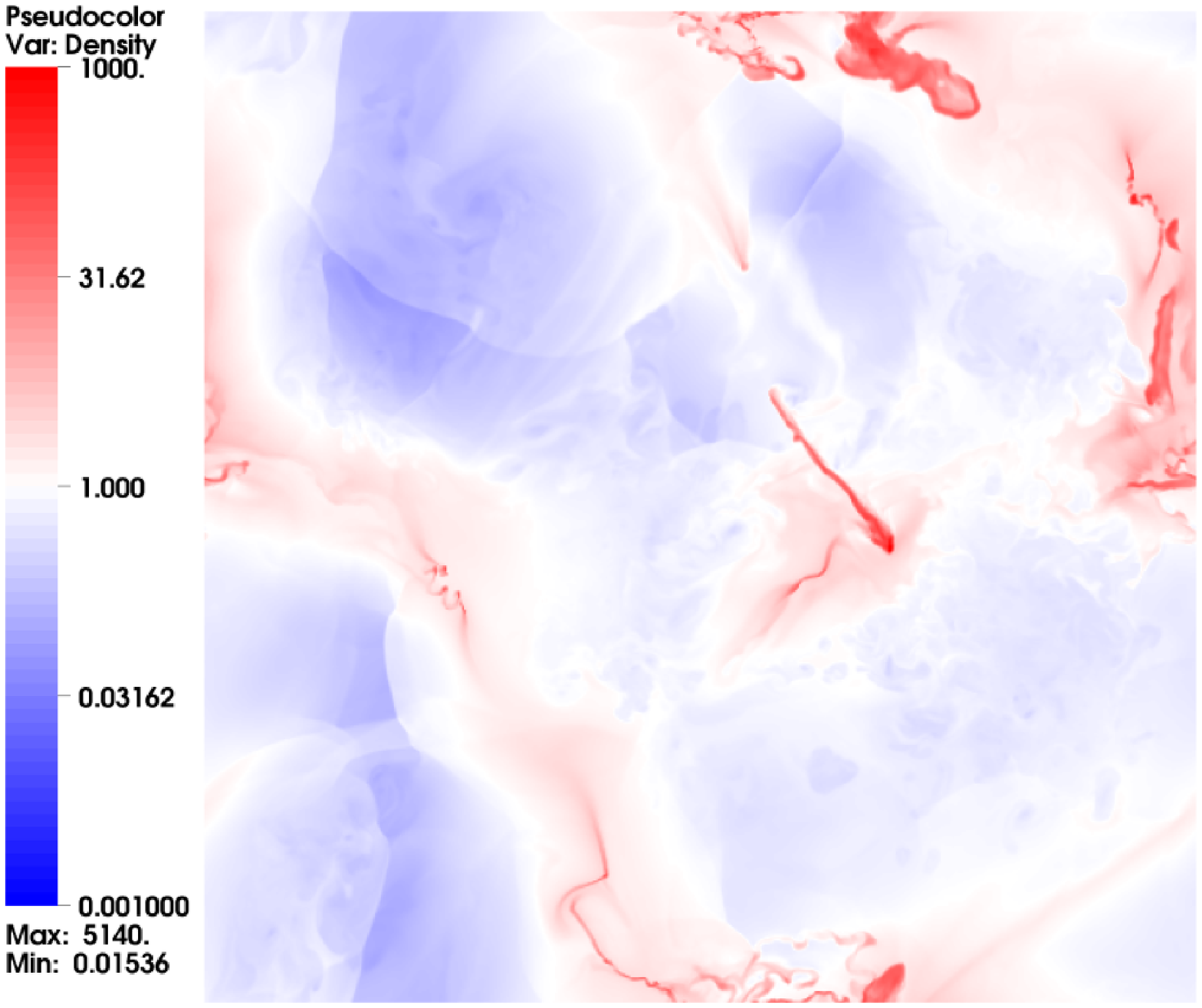}\hspace{12pt}
\includegraphics[width=80mm]{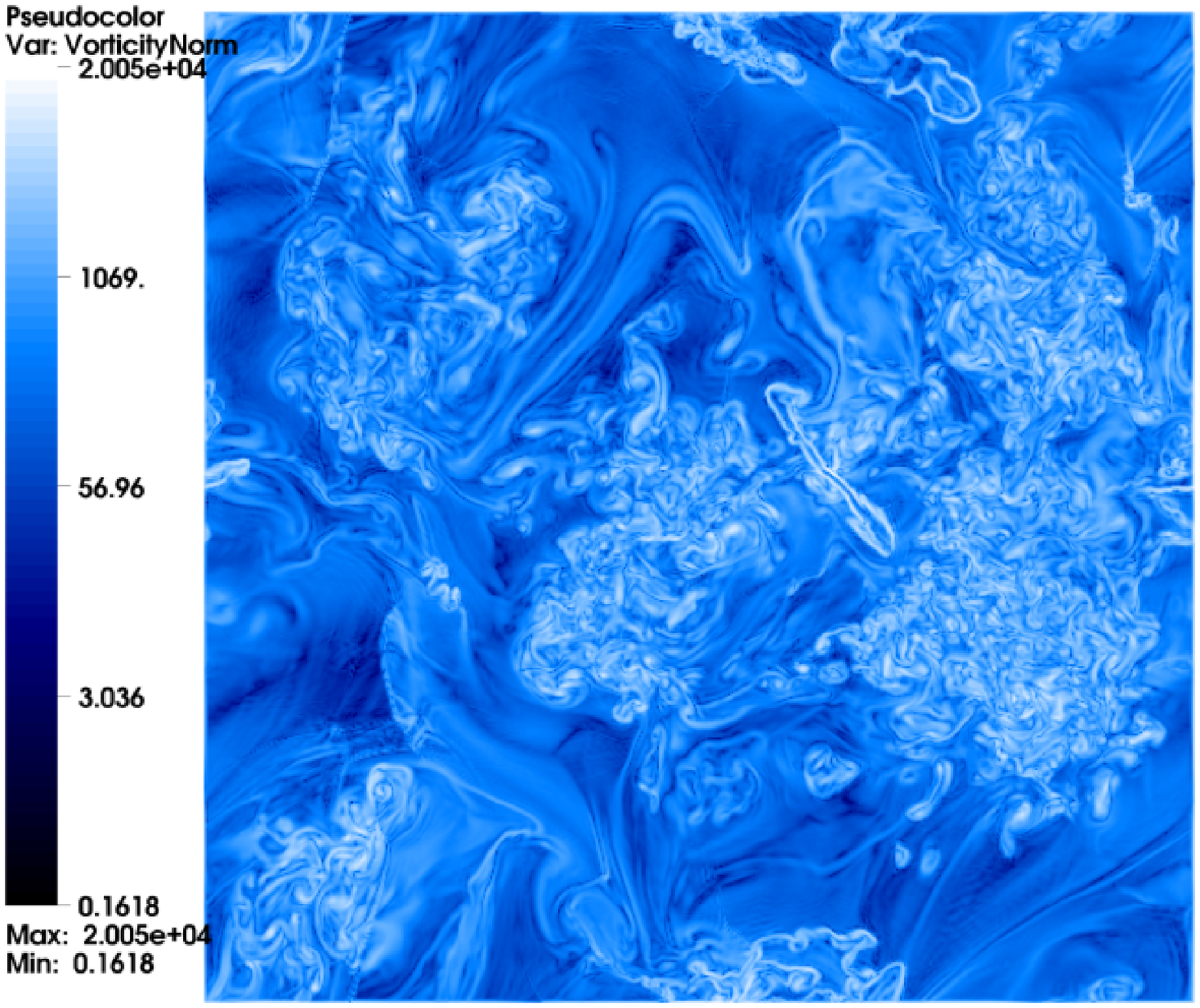}
\caption{Slices of the density (left) and the vorticity field (right) at time $t$ = 1.0 $T$ (top) and 2.0 $T$ (bottom) for run D1Ma1.0\_512.}
\label{fig:7}
\end{figure*}

Slices of the gas density and the vorticity after 1 and 2 integral timescales, which are shown in Fig. \ref{fig:7}, demonstrate that turbulence is very inhomogeneous at these stages of the flow evolution. The patches of high vorticity that can be seen are associated with density fluctuations and shocks in the unstable or warm gas phases. Thus, it appears that the initial production of vorticity mainly results from the forcing. As the TI becomes active and temperature fluctuations grow, this production mechanism is accompanied by baroclinic vorticity generation \citep[see, for instance][]{Schm09}. In the late, nearly stationary phase, the high vorticity is found for the whole range of gas densities (see Fig. \ref{fig:6}). 

\subsection{Influence of the forcing strength}

So far, we have considered a characteristic Mach number \mbox{$\mathrm{Ma} = 1$}, which means that the velocity of the flow produced by the forcing is close to the speed of sound for the initial gas temperature. Next we analyse the influence of varying the strength of compressive forcing for \mbox{$n_0$ = 1.0 cm$^3$} and $N = 256^3$. For the runs D1Ma1.5 and D1Ma2.0, the temperature cutoff at \mbox{30\,000 K} has been applied to avoid artificially high temperatures. In Fig. \ref{fig:8}, the temperature and density PDFs are shown for the characteristic Mach numbers in the range from $0.2$ to $2.0$. In each case, the PDFs are averaged over the last dynamical timescale of the simulation. Except for the temperature PDF of run D1Ma2.0 all other PDFs in the left panel of Fig. \ref{fig:8} show two maxima located in the temperature range of the CNM and the WNM. The positions of the peaks do not vary significantly with the forcing strength in contrast to the peak width. The clearly visible bimodal temperature distribution for the weakly driven simulations D1Ma0.2 and D1Ma0.5 gradually vanishes with increasing forcing strength, while the amount of gas in the unstable range increases. In run D1Ma2.0, no bimodal temperature distribution can be discerned. The density PDFs of the weakly driven simulations with $\mathrm{Ma}<1$ in the right panel of Fig. \ref{fig:8} also show a bimodal shape, where the strong peak in the low-density range ($\sim 0.2\ \mathrm{cm^{-3}}$) and the plateau around \mbox{10 cm$^{-3}$} correspond to high- and low-temperature peaks, respectively. With increasing forcing strength, the maximum of the density PDF broadens substantially. Thus, for high characteristic Mach numbers, the density distribution appears to be increasingly dominated by the nearly adiabatic expansion and compression waves produced by the forcing. Moreover, the high-density tails become flatter; i.~e., stronger forcing tends to induce higher densities. However, even for D1Ma2.0, the peak densities up to several 1000 cm$^{-3}$ are much higher than what is observed in isothermal simulations with comparable characteristic Mach number \citep{Schmidt09}. Consequently, the high-density tails mainly results from the TI \citep[see also][]{Kritsuk02,Li03}. Accordingly, all density PDFs exhibit a power-law like shape in the high-density range, although the numerical resolution is too low to draw definite conclusions.

The pressure-density phase diagrams for D1Ma0.2 and D1Ma1.0, which are shown in Fig. \ref{fig:9}, demonstrate how the thermodynamic state of the gas is affected by turbulence. In the case of turbulence with low intensity (left panel), the distribution shows typical properties of a static two-phase-medium as described by \citet{Field69}. Here the gas exists in a rough pressure equilibrium with deviations from the mean pressure less than half an order of magnitude. Only the high-density gas is out of pressure equilibrium, because of the very small cooling times corresponding to these densities. For stronger turbulent motions, as in run D1Ma1.0 (right panel of Fig. \ref{fig:9}), the gas deviates largely from pressure equilibrium. Thus, for gas in a strongly turbulent state, the classical idea of two distinct phases, coexisting in a rough pressure equilibrium with only small amounts of gas in the unstable range ceases to apply.

To quantify the increasing amount of thermally unstable gas that is observed for strong forcing amplitudes, we determined the mass fractions of the warm, unstable and cold phases as defined in Sect. \ref{sec:simulations}. The results are shown in Fig. \ref{fig:10}, where the different mass fractions are plotted against the total specific kinetic energy for each simulation. In accordance with the PDFs, the mass fractions, as well as the specific kinetic energies, are averaged over the last dynamical timescale. Clearly, the turbulence energy resulting from the forcing has a strong influence on the distribution of the gas into the different phases. The mass fraction of the unstable gas increases at the cost of the mass fraction of the cold gas, whereas the mass fraction of the warm phase is only weakly affected. That the redistribution of the gas with increasing forcing strength mainly occurs between the cold and the unstable phase and, to a lesser extent, between the warm and the unstable phase can be understood as the consequence of the low compressibility of the hot gas, in which only sub- or transonic motions are induced, regardless of the characteristic Mach number. The dynamics in the unstable phase and the cold phase, on the other hand, are highly sensitive to the forcing strength. Although compressive forcing produces the seeds for the TI, we note that high turbulence intensity inhibits the gas from settling into the cold phase. However, it appears that the TI overcomes this effect once a critical amount of cold gas is accumulated. This is suggested by the occurrence of compact regions of cold gas with weakly turbulent interiors (see Fig.~\ref{fig:4}).

\subsection{Tracer particle statistics}
\label{sec:tracers}

To examine the cause of the higher mass fraction of gas in the thermally unstable range for stronger forcing, we included 150\,000 Lagrangian tracer particles in the simulations D1Ma0.2 and D1Ma1.0. This approach is motivated by \citet{Audit05}, who also found an increase in the thermally unstable gas with increasing turbulence intensity at the cost of the cold gas mass. The authors explain this effect on the basis of a simple semi-analytical model. This model implies that the time interval that is spent by a fluid element in the unstable phase before returning to the stable phase becomes longer as the turbulence intensity increases. On the other hand, the mass of gas in the unstable phase is also controlled by the frequency of strong perturbations that bring the gas out of the equilibrium state. Thus, the larger amount of gas in the unstable phase can also be interpreted as a consequence of enhanced turbulent mixing. With the help of the tracer particles, we are able to test the predictions of this model by explicitly following the time evolution of a fluid element moving with the flow. In particular, we measure the time-interval $t_{\mathrm{UP},i}$, defined as the time, a tracer particle is located in the unstable range before returning to the stable state of the CNM or the WNM. We find that the mean duration $\bar{t}_{\mathrm{UP}}$ of the stay of a tracer particle in the unstable phase is reduced by a factor of $2.3$ if the characteristic Mach number of the forcing increases from $0.2$ to $1.0$. However, after analysing how often the particle is perturbed out of the equilibrium state into the unstable phase, we find that the frequency of the perturbations, $\bar{f}_{\mathrm{UP}}$ , rises by a factor of $\sim 4.5$ from run D1Ma0.2 to D1Ma1.0. This implies that the thermally unstable gas fraction roughly increases from run D1Ma0.2 to D1Ma1.0 by a factor of $4.5 \cdot \frac{1}{2.3} \approx 2.0$.

\subsection{Variation in the mean density}

It is also of interest to examine how the gas distribution changes if the mean density of the gas is altered. Interestingly, it turns out that even a small increase of the mean density has a noticeable effect on the distribution of the gas among the different phases. In particular, we performed several simulations with compressive forcing for mean densities of  \mbox{1.8 cm$^{-3}$ and 3.0 cm$^{-3}$} (see Table \ref{table:sim}). Figure \ref{fig:1} shows that the initial state of the gas in these simulations is located closer to the local minimum of the equilibrium curve. For $n_0=3.0\,\mathrm{cm^{-3}}$, the initial state is located near the low-temperature boundary of the central unstable region. The dependence of the mass fractions of the different gas phases on the forcing strength is shown in Fig. \ref{fig:11}. While the qualitative behaviour is similar to what is observed for \mbox{$n_0 = 1.0$ cm$^{-3}$} (see Fig.~\ref{fig:10}), the fraction of cold gas is clearly greater for a higher mean density. In the case $n_0=3.0\ \mathrm{cm^{-3}}$, about $90\ \%$ of the mass settles into the cold phase, more or less independent of the turbulence energy. This result is reasonable because, as the initial state is chosen closer to the cold phase, an increasing amount of the gas will cool and settle into the cold phase due to the external compressions. At densities $n_0\gtrsim 10\ \mathrm{cm^{3}}$, most of the gas resides in the nearly isothermal right branch of the equilibrium curve, and only strong rarefactions can push the gas into the warm phase. In this regime, the dynamical properties are comparable to those of isothermal turbulence. 

\subsection{Solenoidal forcing}

\begin{figure*}
\includegraphics[width=88mm]{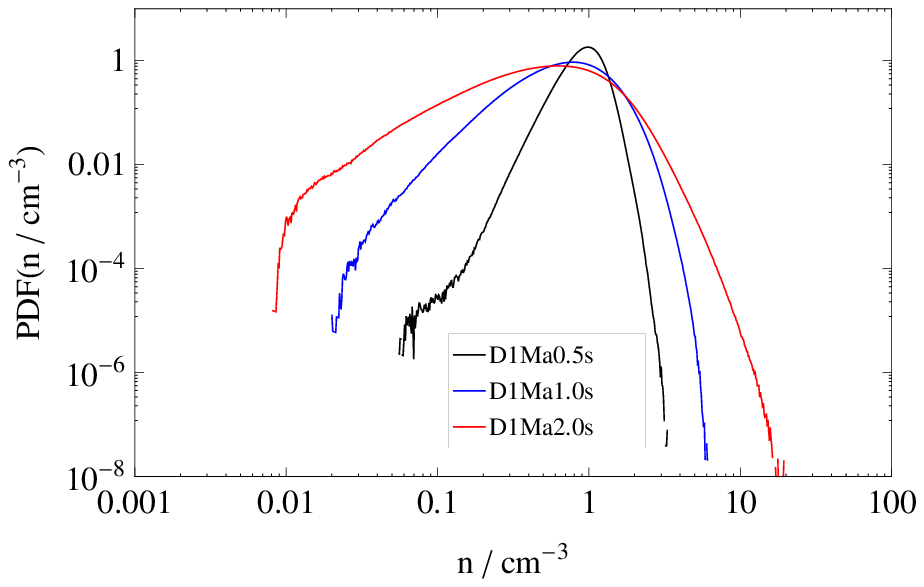}\hspace{12pt}
\includegraphics[width=88mm]{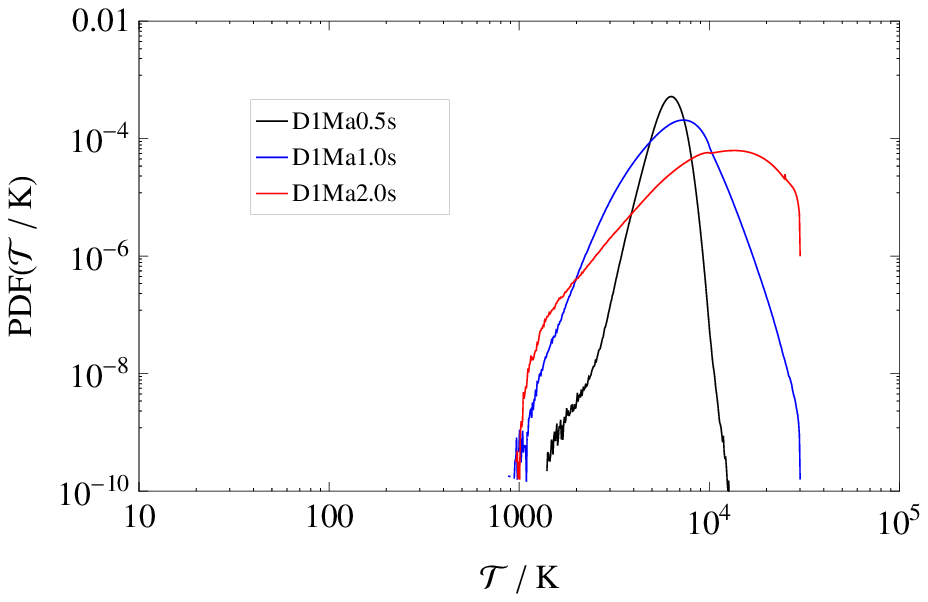}
\caption{Left: Density PDFs of the solenoidally driven runs with mean densities of \mbox{$n_0 = 1.0$ cm$^{-3}$}. Right: Temperature PDFs for the same runs as in the left panel.}
\label{fig:12}
\end{figure*}

\citet{Schmidt09} and \citet{FederDuv09} show in their analysis of supersonic isothermal turbulence that there are significant differences between compressive and solenoidal forcing. In particular, the density PDFs are much broader for compressively driven turbulence \citep{Fed08,FederDuv09}.
For non-isothermal gas, the differences should be even more pronounced because the TI is affected by the different density distributions, which are generated by distinct forms of the forcing. The PDFs of the number density and the temperature resulting from three simulations with solenoidal forcing and a mean density of \mbox{1.0 cm$^{-3}$} (listed in Table \ref{table:sim}) are plotted in Fig. \ref{fig:12}, where the temperature cutoff at \mbox{30\,000 K} was applied for the runs D1Ma1.0s and D1Ma2.0s. As one can see, there is no phase separation at all. Both the density and temperature PDFs show a single peak that roughly corresponds to the initial state. With increasing forcing strength, the distributions become broader, and the gas tends to become hotter. The higher temperatures that are found for \mbox{$\mathrm{Ma} = 2.0$} in comparison to the other runs are caused by the stronger dissipative heating that balances the higher rate of energy injection by the forcing. The amount of cold gas with temperatures below \mbox{200 K} is zero. Thus, it appears that the TI is largely suppressed by solenoidal forcing, although a significant fraction of the gas is in the unstable region (see Fig. \ref{fig:1}). The density PDFs are similar to those reported by \citet{Passot98} for simulations of turbulence in a polytropic gas with $\gamma > 1$, particularly, with regard to the low-density tails. At least for the runs with \mbox{$\mathrm{Ma}\le 1.0$}, there appears to be a power-law range for densities lower than \mbox{1.0 cm$^{-3}$}.

\begin{figure*}
\centering
\includegraphics[width=88mm]{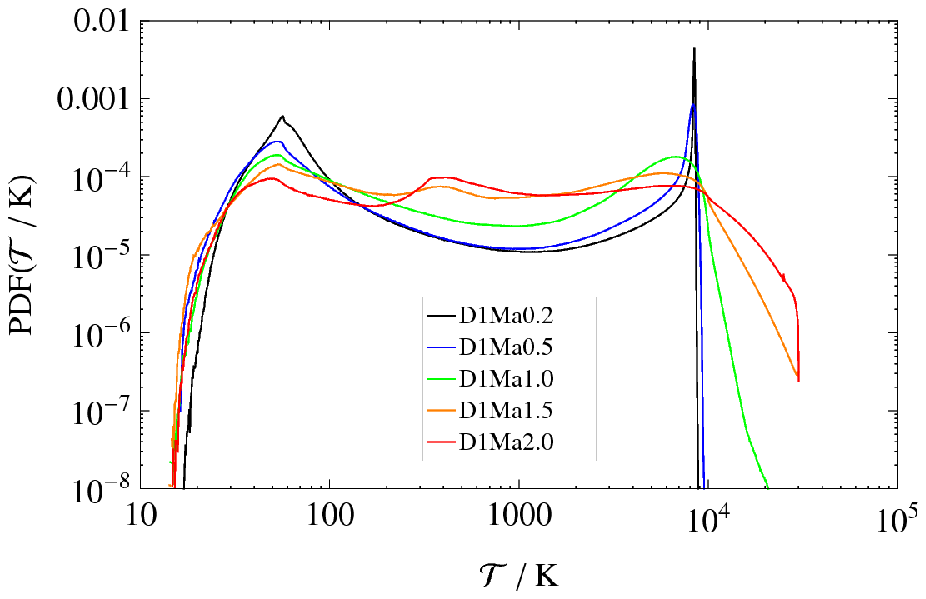}\hspace{12pt}
\includegraphics[width=88mm]{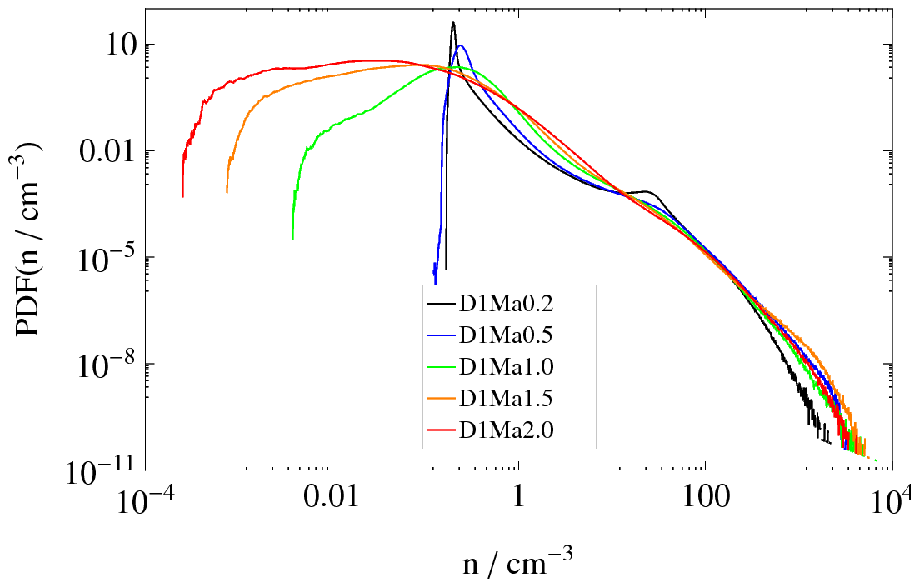}
\caption{Left: Temperature PDFs of simulations with \mbox{$n_0$ = 1.0 cm$^{-3}$} and $N = 256^3$. Except for run D1Ma2.0, the PDFs show two maxima in the warm and cold phase. Right: Density PDFs for the same runs as in the left panel. The PDFs show a peak around \mbox{$n = 0.2$ cm$^{-3}$} and a plateau around \mbox{10 cm$^{-3}$}, which correspond to the WNM and the CNM, respectively.}
\label{fig:8}
\end{figure*}

\begin{figure*}
\centering
\includegraphics[width=88mm]{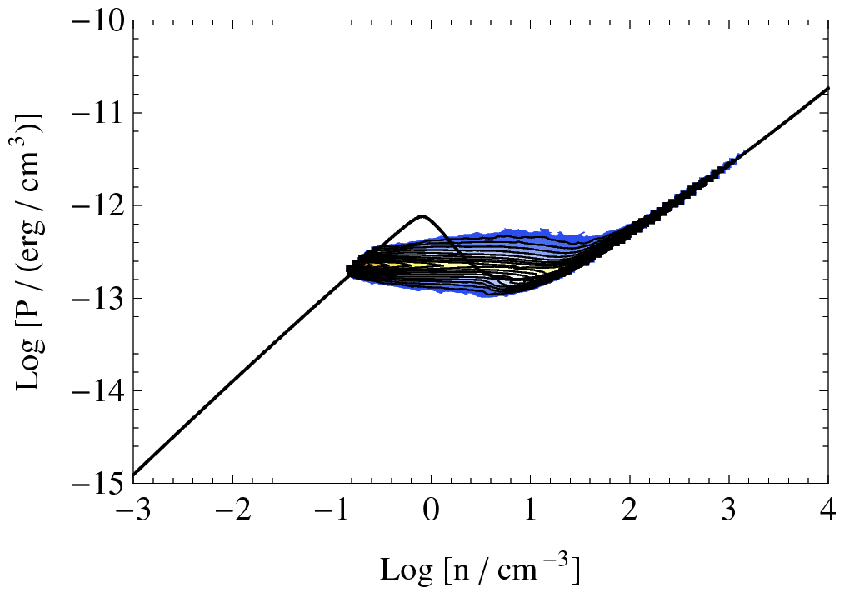}\hspace{12pt}
\includegraphics[width=88mm]{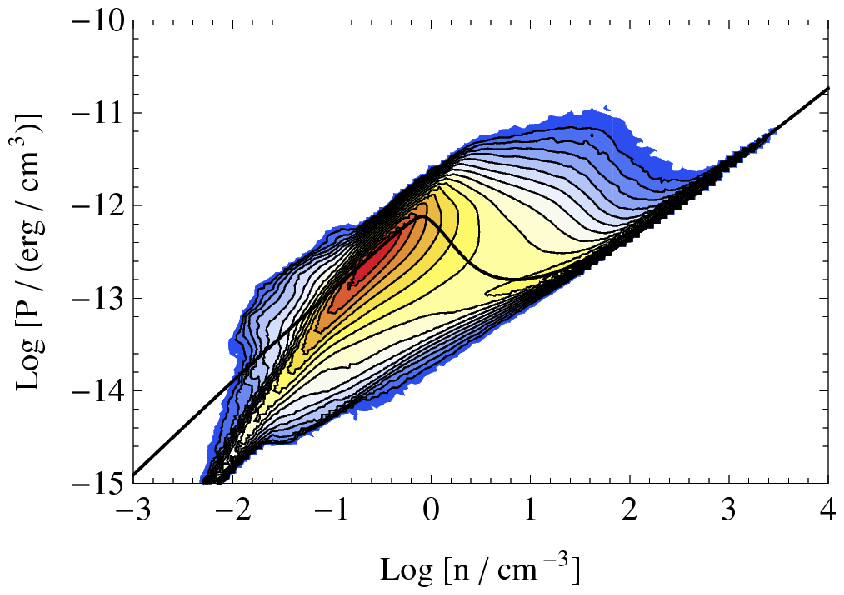}
\caption{Pressure-density phase diagrams of run D1Ma0.2 (left) and D1Ma1.0 (right) with a logarithmic scaling of the contour lines. The equilibrium between heating and cooling is indicated by the solid black line.}
\label{fig:9}
\end{figure*}

\section{Discussion}
\label{sec:discussion}

Most simulations considered in this article were run with a resolution of $N = 256^3$ grid cells. For the chosen box size of $40\ \mathrm{pc}$, the cutoff length in these simulations is $\Delta = 0.16\ \mathrm{pc}$. Consequently, cold gas structures with a typical size in the range of a few parsecs are only marginally resolved and substantially smeared out by numerical diffusion. For this reason, it cannot be expected that the properties of these structures can be described in a detailed quantitative manner. However, the resolution study we performed for a particular parameter set suggests that the turbulent phase separation of the gas can be approximated reasonably well. In particular, we showed that trends can be seen if the forcing is varied, which allows for a qualitative discussion. This will help us selecting parameters for more elaborate numerical studies at higher resolution, using adaptive mesh refinement. In particular, an enhanced numerical resolution will be necessary to confirm the indications for power-law tails of the mass density PDFs toward high densities and to determine the slope with higher precision. These power laws can be found in nearly all simulations of compressively driven turbulence in the present study. This clearly supports the results of \citet{Scalo98}, \citet{Passot98}, and \citet{Li03}, who applied a simple polytropic EOS with $\gamma < 1$ as a simple model for cooling. These authors obtained skewed lognormal density PDFs with power-law tails at high densities. In our simulations, the thermodynamic state of the gas at high densities roughly follows the equilibrium curve, which corresponds to an effective polytropic index $\gamma_{\mathrm{eff}} < 1$. Power-law tails for non-isothermal turbulence followed also from the simulations of decaying turbulence by \citet{Kritsuk02} and from a recent colliding-flow simulation of \citet{AuditHenne09}, in which a polytrope with $\gamma=0.7$ was applied.

\begin{figure}
\resizebox{\hsize}{!}{\includegraphics{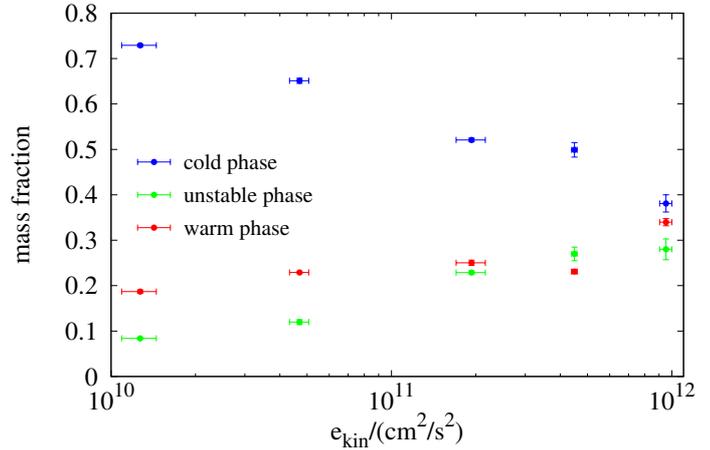}}
\caption{Time-averaged mass fractions of the warm (red points), unstable (green points), and cold phase (blue points) for simulations with \mbox{$n_0 = 1.0$ cm$^{-3}$} and $N = 256^3$ plotted against the specific kinetic energy. An increasing kinetic energy corresponds to an increase in the forcing Mach number (compare  Table \ref{table:sim}).}
\label{fig:10}
\end{figure}

The main result of our study is the increase in the mass of the thermally unstable gas for higher turbulence energy (Fig. \ref{fig:10}), which is roughly balanced by a decrease in the mass of cold gas. Consequently, stronger compressive forcing does not result in more cold, dense gas. The reason is found in the broadening of the pressure distribution with increasing forcing strength and, thus, kinetic energy of the turbulent flow. While weak forcing produces the initial seeds for the TI by perturbing the density of the gas, the forcing drives the gas away from a state of approximate pressure equilibrium if the turbulent velocity becomes comparable to the speed of sound. This effect is shown clearly for runs D1Ma0.2 and D1Ma1.0 in Fig. \ref{fig:9}. Only at high densities is the gas mainly concentrated around the equilibrium curve, independent of forcing strength. This is a consequence of the very short cooling time in comparison to the dynamical timescale and the sound crossing time in this range. At low densities, on the other hand, the gas in run D1Ma1.0 is nearly adiabatic because of the long cooling time in the WNM. The effective polytropic index in this regime is found to be $\gamma_{\mathrm{eff}} = 1.52$. As one can see in Fig. \ref{fig:8} (right panel), this entails a pronounced broadening of the density PDFs in the under-dense range if $\mathrm{Ma}\gtrsim 1$, which is again similar to the results of \citet{Passot98} for their simulations with $\gamma > 1$.

In Sect. \ref{sec:tracers}, we show with the help of tracer particles that the higher frequency of the perturbations of the particles out of the stable regime, $\bar{f}_{\mathrm{UP}}$ ,  overcompensates the decrease in the mean duration,  $\bar{t}_{\mathrm{UP}}$ , a tracer particle stays in the unstable phase if the forcing strength increases. This results in an increase in the unstable mass fraction roughly by a factor of 2.0 for simulation D1Ma1.0 compared to D1Ma0.2, which approximates the ratio of 2.7 that follows from the mass fractions of unstable gas in these simulations reasonably well. Our observation that $\bar{t}_{\mathrm{UP}}$ decreases with increasing forcing strength is contrary to the assumption of \citet{Audit05} that $\bar{t}_{\mathrm{UP}}$ increases with the intensity of turbulence. The discrepancy could appear because their model is 2-dimensional and applies to the unperturbed transition of a fluid element from the WNM to the CNM, which might be insufficient to describe 3-dimensional turbulence with external forcing. In order to rule out that the discrepancy with the model of \citet{Audit05} arises because it only describes transitions from the WNM to the CNM, we also examined phase transitions between the WNM and the CNM with the tracer particles. We found that the mean transition time $\bar{t}_{\mathrm{WNM}\rightarrow\mathrm{CNM}}$ decreases from simulations D1Ma0.2 to D1Ma1.0 by a factor of $\sim 2.5$, which is similar to the behaviour of $\bar{t}_{\mathrm{UP}}$. The averaged time intervals $\bar{t}_{\mathrm{CNM}}$ and $\bar{t}_{\mathrm{WNM}}$, a fluid particle stays in the CNM or in the WNM, become significantly shorter as well. Summarising, the tracer particle statistics imply that phase transitions between the different phases occur more frequently with increasing forcing strength, i.~e., turbulent mixing is more effective, which results in a higher mass fraction of the unstable phase.

\begin{figure*}
\includegraphics[width=88mm]{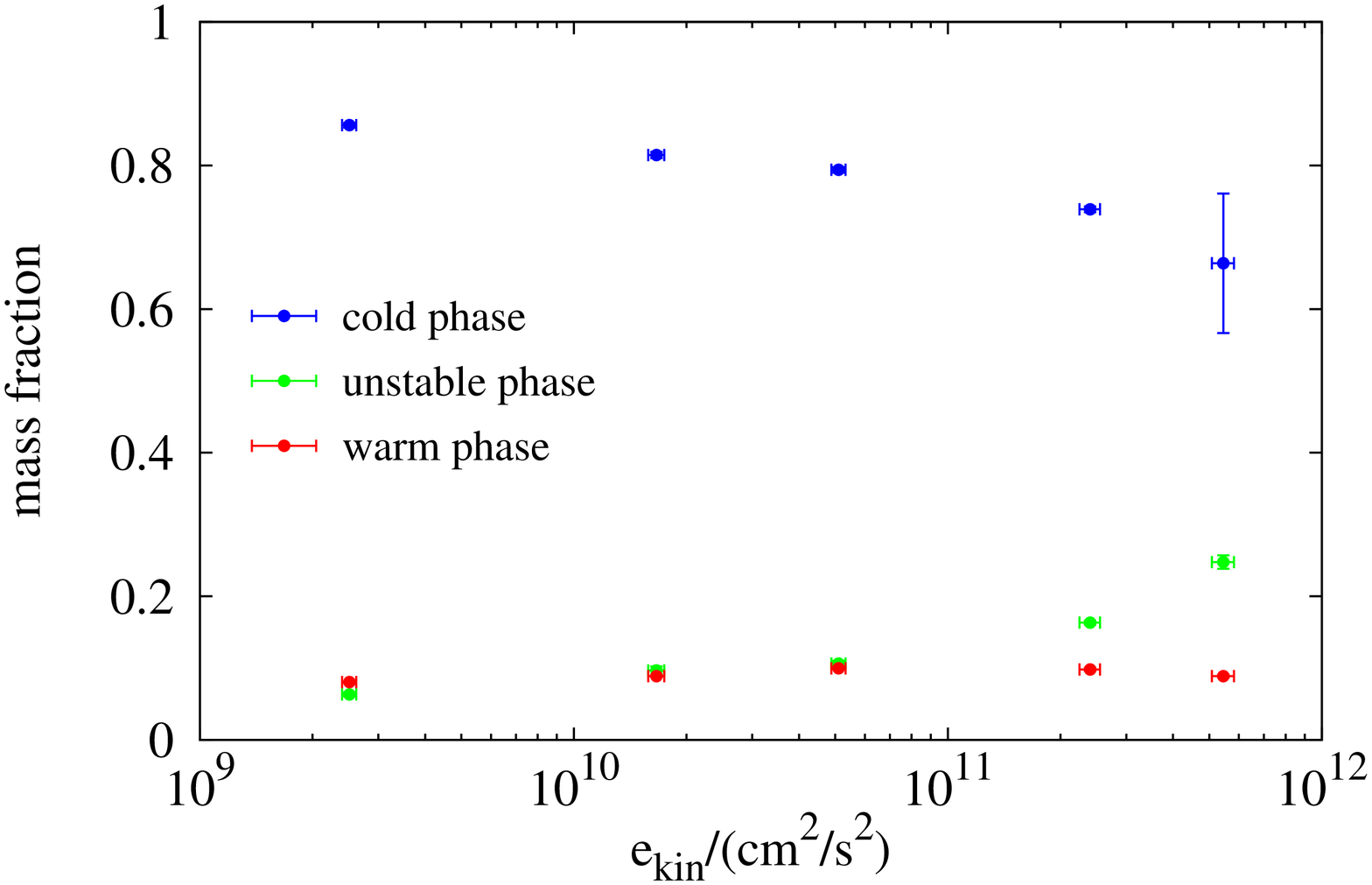}\hspace{12pt}
\includegraphics[width=88mm]{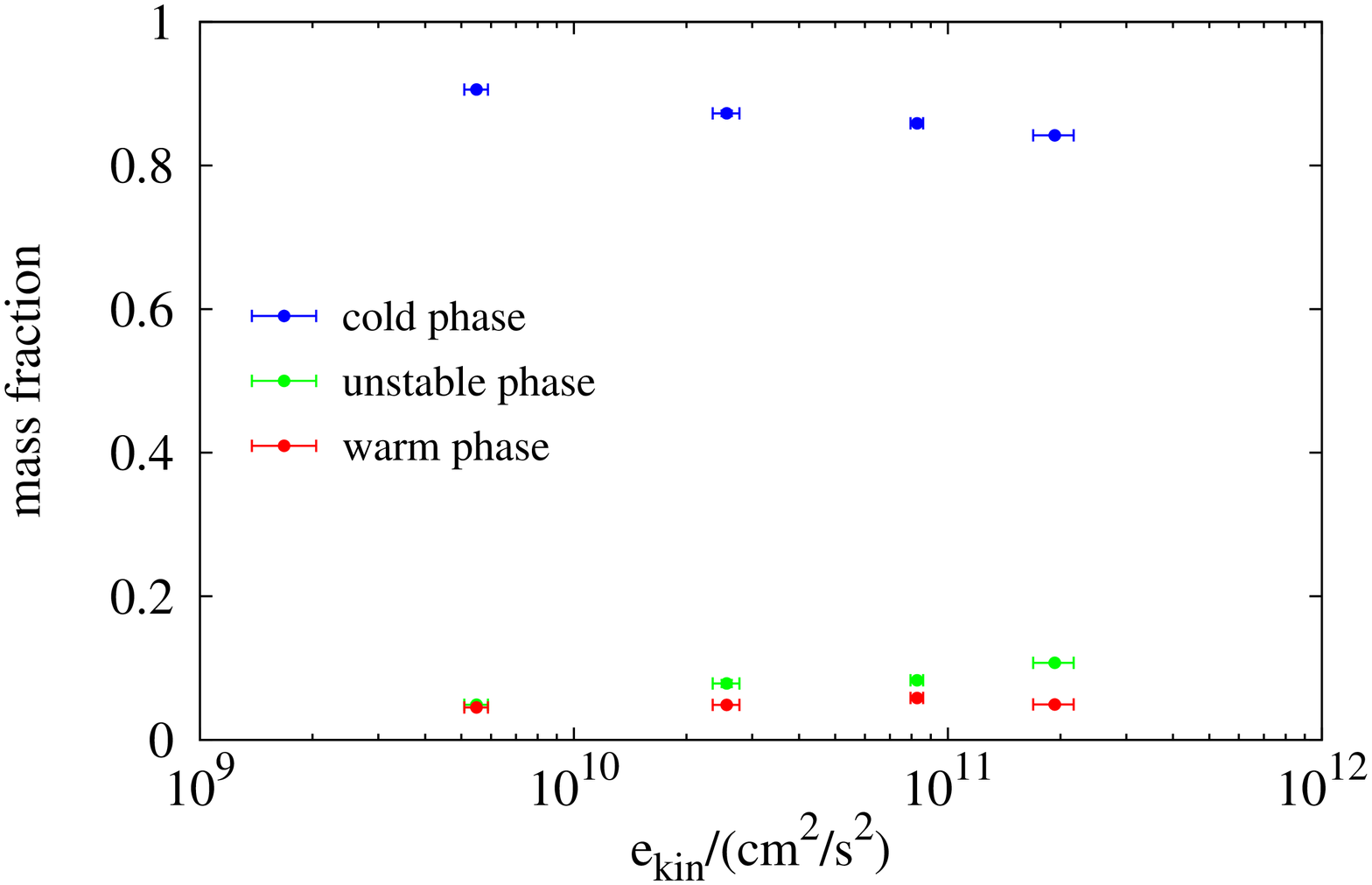}
\caption{Left: Time averaged mass fraction of the warm (red points), unstable (green points) and cold phase (blue points) for the simulations with \mbox{$n_0 = 1.8$ cm$^{-3}$} plotted against the specific kinetic energy. Right: Same as in the left panel but for simulations with \mbox{$n_0 = 3.0$ cm$^{-3}$}.}
\label{fig:11}
\end{figure*}

An important question is for which simulation parameters the mass fractions of the different phases that are plotted in Figs. \ref{fig:10} and \ref{fig:11} are comparable to observational results for the atomic ISM \citep[see, for instance,][]{Heiles01,Heiles03}. One difficulty in such a comparison is that observational values usually do not refer to a particular mean density of the gas, for which measurements are taken. Moreover, the observations presented by these authors do not belong to certain diffuse \ion{H}{i}-clouds but are absorption measurements on cold gas along the lines of sight against different background radio sources. Besides the mean Mach number in the cold phase, which the authors determined to be around 3, particular properties of these objects are not known. Nevertheless, we note that \citet{Heiles03} find that the average mass fraction of atomic gas in the ISM with temperatures in the range from \mbox{500 K} to $5000$ K is at least 29 \%. Together with the warm gas fraction, their measurements imply that the mass fraction of cold gas should be less than 40 \%, which is clearly at the lower end of the range of values following from our numerical study. Thus, strong compressive forcing of turbulence in gas at a mean density of \mbox{1.0 cm$^{-3}$} appears to be typical of the conditions in the diffusive ISM. However, the sources studied by \citet{Heiles03} show a large scatter around the average values, and there exist regions with mass fractions of cold gas up to nearly 90\%, which would correspond to weak forcing and a higher mean density in our study.

A surprising result is that we did not find any cold phase at all in the simulations with purely solenoidal forcing, regardless of the magnitude of the force. Although the turbulent velocity fluctuations in compressible gas entail density fluctuations and the initial state of the gas is in the unstable regime, it appears that the phase separation is completely suppressed. One possible explanation is that the TI would set in on length scales that are smaller than the numerically resolved scales in our simulations.
The minimum length scale of the TI is the Field length \citep{Field65}, which is defined by
\begin{equation}
 \lambda_F = 2 \pi \left[ \frac{\rho}{K}\left( \frac{\rho}{\mathcal{T}} \frac{\partial \fam=2 L}{\partial \rho} - \frac{\partial {\fam=2 L}}{\partial \mathcal{T}} \right) \right]^{-0.5}\,,
\end{equation}
where $K$ is the thermal conductivity. According to \citep{Cox00},
\begin{equation}
 K = 1 \cdot 10^{-6} \, \left(\frac{\mathcal{T}}{1 \mathrm{K}}\right)^{5/2} \mbox{erg cm$^{-1}$ s$^{-1}$ K$^{-1}$.}
\end{equation}
With this, we get a Field length in the CNM that is at least 2 or 3 orders of magnitude smaller than the spatial grid resolution. Therefore, a cold phase might develop at higher resolution. \citet{Hennebelle07a} show in a 2-dimensional resolution study that a spatial resolution below $2 \cdot 10^{-3}$ pc is necessary to reach numerical convergence for a colliding flow of thermally bistable gas, which is unfeasible for 3-dimensional simulations with the current computing power. However, the observed behaviour could also be physical and result from the apparent differences between the turbulent velocity field produced by solenoidal forcing compared to compressively driven turbulence \citep{Schmidt08}. There are indications of an extended range of length scales, for which rotationally driven turbulence has not only lower density fluctuations but also exhibits a more space-filling vorticity field \citep{Fed08,FederDuv09}. As a consequence, the turbulent pressure could stabilise the gas effectively against the TI. An analogous effect has been observed for forced self-gravitating turbulence in an isothermal gas \citep[e.~g.][]{KlessHei00,SchmFeder08c}. In addition, it is likely that the shearing effect of the velocity fluctuations produced by solenoidal forcing disrupts incipient small condensations of dense gas. On the other hand, \citet{Gazol05,Gazol09} and \citet{Kissmann08} found a cold phase for turbulence in non-isothermal gas that was produced by solenoidal forcing. These findings do not necessarily contradict our results, because \citet{Kissmann08} used a mean density of \mbox{2.0 cm$^{-3}$}, and \citet{Gazol05,Gazol09} applied a slightly different cooling function. Thus, the initial state was located deeper inside the unstable regime in both cases. For this reason, the TI can overcome the inhibiting effect of turbulence more easily than in our simulations with $n_{0}=1.0$ cm$^{-3}$. Indeed, we also found a cold phase developing in simulations with solenoidal forcing if the mean density was set to \mbox{3.0 cm$^{-3}$}. Taking these results together, it appears that  the evolution of thermally unstable gas is extremely sensitive to the initial conditions and the details of the cooling function if the forcing is solenoidal.

The strong sensitivity on the mean density and on the forcing shown here could in principle be used to draw some conclusions about the forcing modes in observed diffuse \ion{H}{i}-clouds. In practice, however, it is hard to state something about the underlying forcing because of the difficulty determining the mean gas density observationally, and hence deciding whether the gas is thermally unstable or not. Apart from this difficulty, this work suggests that, for a marginally thermally unstable cloud ($n_{0}\simeq1$ cm$^{-3}$) showing a bimodal temperature distribution, the forcing should mainly be compressive. On the other hand, for clouds in the highly unstable regime ($n_{0}\gtrsim3$ cm$^{-3}$), it is hardly possible to distinguish between solenoidal or compressive forcing as both forcing modes would produce a two-phase-medium with gas distributions that are mostly determined by the thermal instability.

\section{Conclusion}
\label{sec:conclusion}

In this paper, we have performed a parameter study of forced turbulence in thermally bistable gas, with a cooling function that was defined by \citet{Audit05}. Considering both compressive and solenoidal forcing, we find substantial variations in the global statistics and probability density functions of the gas. Because of the limited resolution of the simulations, we did not compute two-point statistics or power spectra, nor did we analyse fragmentation properties. In the following, we summarise our results.
\begin{enumerate}
\item Although we clearly see a phase separation for compressively driven turbulence already at moderate numerical resolution, the PDFs of the mass density and the temperature suggest that a resolution higher than $512^{3}$ is required to accurately determine the tails of the distributions
and the mass fractions of the different phases. For a forcing length scale around $20\ \mathrm{pc}$, a spatial resolution significantly lower than $0.1\ \mathrm{pc}$ is implied.
\item A similarity of the density PDFs of all compressively driven simulations is the power-law like shape in the high-density range, which is caused by an effective polytropic index \mbox{$\gamma_{\mathrm{eff}} < 1$}. Exploring the high-density tails in simulations with enhanced numerical resolution will bear consequences on theoretical predictions for the core mass function \citep{HenneChab09}.
\item For weak forcing, the warm and the unstable phase are nearly isobaric. With increasing forcing strength, we find larger deviations from the pressure equilibrium and a greater amount of gas with temperatures in the intermediate, unstable range, which is roughly balanced by less gas in the cold phase. The Lagrangian statistics inferred from tracer particle suggest that this behaviour is caused by stronger turbulent mixing.
\item The distribution of the gas among the different phases also depends on the mean density. For mean number densities higher than $1.0\ \mathrm{cm^{-3}}$, most of the gas tends to settle into the cold phase.
\item If the forcing is purely solenoidal, we do not find bimodal density and temperature distributions for a mean density of $1.0\ \mathrm{cm^{-3}}$. In this case, all the gas is in the thermally unstable and warm phase.
\end{enumerate}
Qualitatively, our results agree with other numerical studies of turbulence in non-isothermal gas \citep[e.~g.][]{Kritsuk02,Gazol05,Audit05}. The large-scale flow structure produced by compressive stochastic forcing resembles the colliding flow scenario, since transient frontal collisions of shocks occur randomly \citep{Schmidt09}. Given the sensitivity of the phase separation on the mean density, a crucial difference arises from the accumulation of mass in the colliding flow scenario, because of the inflow boundary conditions. In a periodic box, on the other hand, the mass is conserved. Consequently, in colliding flow simulations, there is an implicit drift through the parameter space in the course of time. In this regard, we note that \citet{AuditHenne09} do not find clear indications of a power-law tail of the density PDF in their simulations with cooling. If self-gravity is included, a caveat of periodic-box simulations is that the amount of mass that is confined in gravitationally collapsing objects cannot  indefinitely increase at a constant rate. A net inflow of gas from the boundaries, on the other hand, constantly rejuvenates the supply of gas that can be bound by gravitational collapse \citep[e.~g.][]{Vazquez07,HenBan08,Banerjee09}, although a statistically stationary state cannot be reached either because of the reasons mentioned above. Whether the physical conditions in the interstellar medium on a length scale of a few 10 pc are approximated more closely in simulations of colliding flows without external forcing compared to forced turbulence simulations at a fixed mean density remains a matter of debate. Regardless of this, the results presented here show a substantial sensitivity on the initial state of the gas and the forcing properties. This illustrates that a restriction to highly specialised conditions possibly leads to misleading conclusions, as even minor changes in the studied conditions can cause significantly different results.

In large-scale simulations of the ISM \citep[e.~g.][]{Slyz05,Joung06,Joung09}, the spatial resolutions are usually above \mbox{1 pc}, and in simulations of galactic disks, the resolution is even coarser \citep[e.~g.][]{AgerLake09,TaskTan09}. Thus, as shown here, it is quite likely that the mass fractions of the gas in the different phases would change significantly if smaller length scales could be resolved, particularly in the cold phase and in the low-temperature segment of the unstable phase. This has a potential impact on calculations of the star formation rate, because only cold gas can turn into stars. A subgrid-model for star formation that takes the multi-phase structure of the ISM into account has already been developed and applied by \citet{Springel03}, although these authors omitted the influence of turbulent motions that we focussed on. However, our results indicate that the errors due to unresolved thermodynamic processes will be small if the gas density in the grid cells where sink particles representing star-forming clouds are added is much higher than $1.0\ \mathrm{cm^{-3}}$. Only in regions that just begin to contract, a correction of the star formation efficiency by the mass fraction of cold gas might slow down the growth of the stellar mass in disk-galaxy simulations. In the dense clumps that are formed subsequently, however, most of the gas would cool down to molecular-cloud temperatures if a sufficiently low temperature floor was applied, and it can be assumed that the local star formation efficiency is regulated mainly by gravity and small-scale turbulence \citep{KrumholzMcKee05,PadNord09,KlessHenne09}. 

For further studies of turbulence in the multi-phase ISM, adaptive mesh refinement (AMR) is one method to increase the numerical resolution in such simulations. Although \citet{Niklaus09} show that there are serious difficulties in reproducing the properties of 2-dimensional colliding flows by means of AMR even if sophisticated refinement criteria are applied, we expect that the situation will be more favourable to 3-dimensional simulations of forced turbulence. If the low enstrophy (i.~e., the volume-integrated square of the vorticity) of compact cold gas regions could be confirmed in AMR simulations, it would point to an additional driving mechanism to sustain supersonic turbulence in molecular cloud interiors. Examples include external driving by gravitational accretion \citep[e.~g.][]{KlessHenne09} or internal driving by massive stars \citep[e.~g.][]{FallKrum09}. Moreover, there is also room for methodological improvements in the cooling. Firstly, the range of the cooling function should be extended to temperatures higher than \mbox{10\,000 K}. On the other hand, the cooling of the gas at low temperatures will be altered if chemical processes, in particular, the generation and destruction of molecular hydrogen, are explicitly included \citep{GlovFed09}. Secondly, it is not clear whether an explicit scheme with subcycling will be sufficient to treat the source terms in the energy equation if the effective numerical resolution becomes very high. For subsequent simulations it would also be interesting to study the effect of magnetic fields.

\begin{acknowledgements}
We thank Edouard Audit and Patrick Hennebelle for providing their cooling function. We also thank Christoph Federrath for providing his tracer particle algorithm. The numerical simulations presented in this article were performed on HLRB2 at the Leibniz Supercomputing Centre in Garching, as part of project h0972 using the Enzo code developed by the Laboratory for Computational Astrophysics at the University of California in San Diego (http://lca.ucsd.edu). D. Seifried is currently founded by the DFG via the Emmy-Noether grant BA3706.
\end{acknowledgements}

% \bibliographystyle{aa}
% \bibliography{14373lit}

\begin{thebibliography}{64}
\expandafter\ifx\csname natexlab\endcsname\relax\def\natexlab#1{#1}\fi

\bibitem[{{Agertz} {et~al.}(2009){Agertz}, {Lake}, {Teyssier}, {Moore},
  {Mayer}, \& {Romeo}}]{AgerLake09}
{Agertz}, O., {Lake}, G., {Teyssier}, R., {et~al.} 2009, \mnras, 392, 294

\bibitem[{{Audit} \& {Hennebelle}(2005)}]{Audit05}
{Audit}, E. \& {Hennebelle}, P. 2005, \aap, 433, 1

\bibitem[{{Audit} \& {Hennebelle}(2010)}]{AuditHenne09}
{Audit}, E. \& {Hennebelle}, P. 2010, \aap, 511, A76+

\bibitem[{{Bakes} \& {Tielens}(1994)}]{Bakes94}
{Bakes}, E.~L.~O. \& {Tielens}, A.~G.~G.~M. 1994, \apj, 427, 822

\bibitem[{{Ballesteros-Paredes} {et~al.}(2007){Ballesteros-Paredes}, {Klessen},
  {Mac Low}, \& {Vazquez-Semadeni}}]{BallKless07}
{Ballesteros-Paredes}, J., {Klessen}, R.~S., {Mac Low}, M.-M., \&
  {Vazquez-Semadeni}, E. 2007, in Protostars and Planets V, ed. B.~{Reipurth},
  D.~{Jewitt}, \& K.~{Keil}, 63--80

\bibitem[{{Banerjee} {et~al.}(2009){Banerjee}, {V{\'a}zquez-Semadeni},
  {Hennebelle}, \& {Klessen}}]{Banerjee09}
{Banerjee}, R., {V{\'a}zquez-Semadeni}, E., {Hennebelle}, P., \& {Klessen},
  R.~S. 2009, \mnras, 398, 1082

\bibitem[{{Colella} \& {Woodward}(1984)}]{Colella84}
{Colella}, P. \& {Woodward}, P.~R. 1984, Journal of Computational Physics, 54,
  174

\bibitem[{{Cox}(2000)}]{Cox00}
{Cox}, A.~N. 2000, Allen's Astrophysical Quantities, 4th edn. (Springer-Verlag)

\bibitem[{{Dickey} {et~al.}(1977){Dickey}, {Salpeter}, \& {Terzian}}]{Dickey77}
{Dickey}, J.~M., {Salpeter}, E.~E., \& {Terzian}, Y. 1977, \apjl, 211, L77+

\bibitem[{{Dobbs} {et~al.}(2008){Dobbs}, {Glover}, {Clark}, \&
  {Klessen}}]{DobbsGlov08}
{Dobbs}, C.~L., {Glover}, S.~C.~O., {Clark}, P.~C., \& {Klessen}, R.~S. 2008,
  \mnras, 389, 1097

\bibitem[{{Elmegreen} \& {Scalo}(2004)}]{ElmeScalo04}
{Elmegreen}, B.~G. \& {Scalo}, J. 2004, \araa, 42, 211

\bibitem[{{Eswaran} \& {Pope}(1988)}]{Eswaran88}
{Eswaran}, V. \& {Pope}, S.~B. 1988, Computers and Fluids, 16, 257

\bibitem[{{Fall} {et~al.}(2010){Fall}, {Krumholz}, \& {Matzner}}]{FallKrum09}
{Fall}, S.~M., {Krumholz}, M.~R., \& {Matzner}, C.~D. 2010, \apjl, 710, L142

\bibitem[{{Federrath} {et~al.}(2009{\natexlab{a}}){Federrath}, {Duval},
  {Klessen}, {Schmidt}, \& {Mac Low}}]{FederDuv09}
{Federrath}, C., {Duval}, J., {Klessen}, R., {Schmidt}, W., \& {Mac Low}, M.~M.
  2009{\natexlab{a}}, {e-print arXiv:0905.1060}, accepted for publication by
  \aap

\bibitem[{{Federrath} {et~al.}(2008){Federrath}, {Klessen}, \&
  {Schmidt}}]{Fed08}
{Federrath}, C., {Klessen}, R.~S., \& {Schmidt}, W. 2008, \apjl, 688, L79

\bibitem[{{Federrath} {et~al.}(2009{\natexlab{b}}){Federrath}, {Klessen}, \&
  {Schmidt}}]{Fed09}
{Federrath}, C., {Klessen}, R.~S., \& {Schmidt}, W. 2009{\natexlab{b}}, \apj,
  692, 364

\bibitem[{{Field}(1965)}]{Field65}
{Field}, G.~B. 1965, \apj, 142, 531

\bibitem[{{Field} {et~al.}(1969){Field}, {Goldsmith}, \& {Habing}}]{Field69}
{Field}, G.~B., {Goldsmith}, D.~W., \& {Habing}, H.~J. 1969, \apjl, 155, L149+

\bibitem[{{Folini} {et~al.}(2009){Folini}, {Walder}, \& {Favre}}]{FoliniWald09}
{Folini}, D., {Walder}, R., \& {Favre}, J.~M. 2009, e-prints arXiv:0912.2496

\bibitem[{{Gazol} {et~al.}(2009){Gazol}, {Luis}, \& {Kim}}]{Gazol09}
{Gazol}, A., {Luis}, L., \& {Kim}, J. 2009, \apj, 693, 656

\bibitem[{{Gazol} {et~al.}(2005){Gazol}, {V{\'a}zquez-Semadeni}, \&
  {Kim}}]{Gazol05}
{Gazol}, A., {V{\'a}zquez-Semadeni}, E., \& {Kim}, J. 2005, \apj, 630, 911

\bibitem[{{Glover} {et~al.}(2010){Glover}, {Federrath}, {Mac Low}, \&
  {Klessen}}]{GlovFed09}
{Glover}, S.~C.~O., {Federrath}, C., {Mac Low}, M., \& {Klessen}, R.~S. 2010,
  \mnras, 404, 2

\bibitem[{{Heiles}(2001)}]{Heiles01}
{Heiles}, C. 2001, \apjl, 551, L105

\bibitem[{{Heiles} \& {Troland}(2003)}]{Heiles03}
{Heiles}, C. \& {Troland}, T.~H. 2003, \apj, 586, 1067

\bibitem[{{Heitsch} {et~al.}(2006){Heitsch}, {Slyz}, {Devriendt}, {Hartmann},
  \& {Burkert}}]{Heitsch06}
{Heitsch}, F., {Slyz}, A.~D., {Devriendt}, J.~E.~G., {Hartmann}, L.~W., \&
  {Burkert}, A. 2006, \apj, 648, 1052

\bibitem[{{Hennebelle} \& {Audit}(2007)}]{Hennebelle07a}
{Hennebelle}, P. \& {Audit}, E. 2007, \aap, 465, 431

\bibitem[{{Hennebelle} {et~al.}(2008){Hennebelle}, {Banerjee},
  {V{\'a}zquez-Semadeni}, {Klessen}, \& {Audit}}]{HenBan08}
{Hennebelle}, P., {Banerjee}, R., {V{\'a}zquez-Semadeni}, E., {Klessen}, R.~S.,
  \& {Audit}, E. 2008, \aap, 486, L43

\bibitem[{{Hennebelle} \& {Chabrier}(2009)}]{HenneChab09}
{Hennebelle}, P. \& {Chabrier}, G. 2009, \apj, 702, 1428

\bibitem[{{Hunter}(1970)}]{Hunter70}
{Hunter}, Jr., J.~H. 1970, \apj, 161, 451

\bibitem[{{Joung} \& {Mac Low}(2006)}]{Joung06}
{Joung}, M.~K.~R. \& {Mac Low}, M.-M. 2006, \apj, 653, 1266

\bibitem[{{Joung} {et~al.}(2009){Joung}, {Mac Low}, \& {Bryan}}]{Joung09}
{Joung}, M.~R., {Mac Low}, M.-M., \& {Bryan}, G.~L. 2009, \apj, 704, 137

\bibitem[{{Kanekar} {et~al.}(2003){Kanekar}, {Subrahmanyan}, {Chengalur}, \&
  {Safouris}}]{Kanekar03}
{Kanekar}, N., {Subrahmanyan}, R., {Chengalur}, J.~N., \& {Safouris}, V. 2003,
  \mnras, 346, L57

\bibitem[{{Kissmann} {et~al.}(2008){Kissmann}, {Kleimann}, {Fichtner}, \&
  {Grauer}}]{Kissmann08}
{Kissmann}, R., {Kleimann}, J., {Fichtner}, H., \& {Grauer}, R. 2008, \mnras,
  391, 1577

\bibitem[{{Klessen} {et~al.}(2000){Klessen}, {Heitsch}, \& {Mac
  Low}}]{KlessHei00}
{Klessen}, R.~S., {Heitsch}, F., \& {Mac Low}, M. 2000, \apj, 535, 887

\bibitem[{{Klessen} \& {Hennebelle}(2010)}]{KlessHenne09}
{Klessen}, R.~S. \& {Hennebelle}, P. 2010, \aap, 520, A17+

\bibitem[{{Kritsuk} \& {Norman}(2002)}]{Kritsuk02}
{Kritsuk}, A.~G. \& {Norman}, M.~L. 2002, \apjl, 569, L127

\bibitem[{{Kritsuk} {et~al.}(2007){Kritsuk}, {Norman}, {Padoan}, \&
  {Wagner}}]{Kritsuk07}
{Kritsuk}, A.~G., {Norman}, M.~L., {Padoan}, P., \& {Wagner}, R. 2007, \apj,
  665, 416

\bibitem[{{Krumholz} \& {McKee}(2005)}]{KrumholzMcKee05}
{Krumholz}, M.~R. \& {McKee}, C.~F. 2005, \apj, 630, 250

\bibitem[{{Li} {et~al.}(2003){Li}, {Klessen}, \& {Mac Low}}]{Li03}
{Li}, Y., {Klessen}, R.~S., \& {Mac Low}, M.-M. 2003, \apj, 592, 975

\bibitem[{{Mac Low} \& {Klessen}(2004)}]{MacLow04}
{Mac Low}, M. \& {Klessen}, R.~S. 2004, Reviews of Modern Physics, 76, 125

\bibitem[{{Mac Low} {et~al.}(1998){Mac Low}, {Klessen}, {Burkert}, \&
  {Smith}}]{MacLow98}
{Mac Low}, M.-M., {Klessen}, R.~S., {Burkert}, A., \& {Smith}, M.~D. 1998,
  Physical Review Letters, 80, 2754

\bibitem[{{McKee} \& {Ostriker}(2007)}]{KeeOst07}
{McKee}, C.~F. \& {Ostriker}, E.~C. 2007, \araa, 45, 565

\bibitem[{{McKee} \& {Ostriker}(1977)}]{MO77}
{McKee}, C.~F. \& {Ostriker}, J.~P. 1977, \apj, 218, 148

\bibitem[{{Niklaus} {et~al.}(2009){Niklaus}, {Schmidt}, \&
  {Niemeyer}}]{Niklaus09}
{Niklaus}, M., {Schmidt}, W., \& {Niemeyer}, J.~C. 2009, \aap, 506, 1065

\bibitem[{{O'Shea} {et~al.}(2005){O'Shea}, {Bryan}, {Bordner}, {Norman},
  {Abel}, {Harkness}, \& {Kritsuk}}]{OShea05}
{O'Shea}, B.~W., {Bryan}, G., {Bordner}, J., {et~al.} 2005, in Springer Lecture
  Notes in Computational Science and Engineering, Vol.~41, Adaptive Mesh
  Refinement - Theory and Applications, ed. T.~{Plewa}, T.~{Linde}, \& V.~G.
  {Weirs}

\bibitem[{{Padoan} \& {Nordlund}(2009)}]{PadNord09}
{Padoan}, P. \& {Nordlund}, A. 2009, e-print arXiv:0907.0248

\bibitem[{{Passot} \& {V{\'a}zquez-Semadeni}(1998)}]{Passot98}
{Passot}, T. \& {V{\'a}zquez-Semadeni}, E. 1998, \pre, 58, 4501

\bibitem[{{Scalo} {et~al.}(1998){Scalo}, {Vazquez-Semadeni}, {Chappell}, \&
  {Passot}}]{Scalo98}
{Scalo}, J., {Vazquez-Semadeni}, E., {Chappell}, D., \& {Passot}, T. 1998,
  \apj, 504, 835

\bibitem[{{Schmidt}(2007)}]{Schm09}
{Schmidt}, W. 2007, in {Structure Formation in the Universe}, ed. G.~{Chabrier}
  ({Cambridge Contemporary Astrophysics, Cambridge University Press}), 20--36

\bibitem[{{Schmidt} {et~al.}(2009){Schmidt}, {Federrath}, {Hupp}, {Kern}, \&
  {Niemeyer}}]{Schmidt09}
{Schmidt}, W., {Federrath}, C., {Hupp}, M., {Kern}, S., \& {Niemeyer}, J.~C.
  2009, \aap, 494, 127

\bibitem[{{Schmidt} {et~al.}(2008{\natexlab{a}}){Schmidt}, {Federrath}, {Hupp},
  {Maier}, \& {Niemeyer}}]{SchmFeder08c}
{Schmidt}, W., {Federrath}, C., {Hupp}, M., {Maier}, A., \& {Niemeyer}, J.~C.
  2008{\natexlab{a}}, in High Performance Computing in Science and Engineering,
  Garching/Munich 2007, ed. S.~{Wagner}, M.~{Steinmetz}, A.~{Bode}, \&
  M.~{Brehm} (Springer)

\bibitem[{{Schmidt} {et~al.}(2008{\natexlab{b}}){Schmidt}, {Federrath}, \&
  {Klessen}}]{Schmidt08}
{Schmidt}, W., {Federrath}, C., \& {Klessen}, R. 2008{\natexlab{b}}, Physical
  Review Letters, 101, 194505

\bibitem[{{Schmidt} {et~al.}(2006){Schmidt}, {Hillebrandt}, \&
  {Niemeyer}}]{Schmidt06}
{Schmidt}, W., {Hillebrandt}, W., \& {Niemeyer}, J.~C. 2006, Computers and
  Fluids, 35, 353

\bibitem[{{Slyz} {et~al.}(2005){Slyz}, {Devriendt}, {Bryan}, \&
  {Silk}}]{Slyz05}
{Slyz}, A.~D., {Devriendt}, J.~E.~G., {Bryan}, G., \& {Silk}, J. 2005, \mnras,
  356, 737

\bibitem[{{Spitzer}(1978)}]{Spitzer78}
{Spitzer}, L. 1978, Physical Processes in the Interstellar Medium, 1st edn.
  (Wiley-Interscience)

\bibitem[{{Spitzer} \& {Fitzpatrick}(1995)}]{Spitzer95}
{Spitzer}, Jr., L. \& {Fitzpatrick}, E.~L. 1995, \apj, 445, 196

\bibitem[{{Springel} \& {Hernquist}(2003)}]{Springel03}
{Springel}, V. \& {Hernquist}, L. 2003, \mnras, 339, 289

\bibitem[{{Stone} {et~al.}(1998){Stone}, {Ostriker}, \& {Gammie}}]{Stone98}
{Stone}, J.~M., {Ostriker}, E.~C., \& {Gammie}, C.~F. 1998, \apjl, 508, L99

\bibitem[{{Tasker} \& {Bryan}(2006)}]{TaskBry06}
{Tasker}, E.~J. \& {Bryan}, G.~L. 2006, \apj, 641, 878

\bibitem[{{Tasker} \& {Tan}(2009)}]{TaskTan09}
{Tasker}, E.~J. \& {Tan}, J.~C. 2009, \apj, 700, 358

\bibitem[{{V{\'a}zquez-Semadeni} {et~al.}(2003){V{\'a}zquez-Semadeni}, {Gazol},
  {Passot}, \& {et al.}}]{Vazquez03}
{V{\'a}zquez-Semadeni}, E., {Gazol}, A., {Passot}, T., \& {et al.} 2003, in
  Lecture Notes in Physics, Berlin Springer Verlag, Vol. 614, Turbulence and
  Magnetic Fields in Astrophysics, ed. {E.~Falgarone \& T.~Passot}, 213--251

\bibitem[{{V{\'a}zquez-Semadeni} {et~al.}(2007){V{\'a}zquez-Semadeni},
  {G{\'o}mez}, {Jappsen}, {Ballesteros-Paredes}, {Gonz{\'a}lez}, \&
  {Klessen}}]{Vazquez07}
{V{\'a}zquez-Semadeni}, E., {G{\'o}mez}, G.~C., {Jappsen}, A.~K., {et~al.}
  2007, \apj, 657, 870

\bibitem[{{Wolfire} {et~al.}(1995){Wolfire}, {Hollenbach}, {McKee}, {Tielens},
  \& {Bakes}}]{Wolfire95}
{Wolfire}, M.~G., {Hollenbach}, D., {McKee}, C.~F., {Tielens}, A.~G.~G.~M., \&
  {Bakes}, E.~L.~O. 1995, \apj, 443, 152

\bibitem[{{Wolfire} {et~al.}(2003){Wolfire}, {McKee}, {Hollenbach}, \&
  {Tielens}}]{Wolfire03}
{Wolfire}, M.~G., {McKee}, C.~F., {Hollenbach}, D., \& {Tielens}, A.~G.~G.~M.
  2003, \apj, 587, 278

\end{thebibliography}

\end{document}